\documentclass[%
 reprint,
superscriptaddress,
nofootinbib,
amsmath,amssymb,
aps,
onecolumn,
]{revtex4-2}

\usepackage{dsfont}
\usepackage{physics}
\usepackage{mathrsfs}
\usepackage{amsmath}
\usepackage{indentfirst}
\usepackage{amssymb}
\usepackage{color}
\usepackage{amsfonts} 
\usepackage[framemethod=TikZ]{mdframed}
\usepackage{mathtools}
\usepackage[retainorgcmds]{IEEEtrantools}
\usepackage[colorlinks=true,linkcolor=blue,urlcolor=blue,citecolor=blue,pdfusetitle]{hyperref}
\usepackage{times,txfonts}
\usepackage{graphicx}
\usepackage{dcolumn}
\usepackage{bm}

\usepackage[colorlinks=true,linkcolor=blue,urlcolor=blue,citecolor=blue,pdfusetitle]{hyperref}
\usepackage{hyperref}
\begin{document}

\title{HARNESSING Nth ROOT GATES FOR ENERGY STORAGE}
\author{Elliot Fox}
\email{elliot.fox@york.ac.uk}
\affiliation{School of Physics, Engineering and Technology, University of York, York YO10 5DD, United Kingdom}
\author{Marcela Herrera}
\email{amherrera@uao.edu.co}
\affiliation{Facultad de Ingeniería y Ciencias Básicas, Universidad Autónoma de Occidente, Cali 760030}
\author{Ferdinand Schmidt-Kaler}
\email{fsk@uni-mainz.de}
\affiliation{Quantum, Institut für Physik, Universität Mainz, D-55128 Mainz, Germany}
\author{Irene D'Amico}
\email{irene.damico@york.ac.uk}
\affiliation{School of Physics, Engineering and Technology, University of York, York YO10 5DD, United Kingdom}
\date{\today}

\begin{abstract}We explore the use of fractional control-not gates in  quantum thermodynamics. The Nth-root gate allows for a paced application of two-qubit operations. We apply it in quantum thermodynamic protocols for charging a quantum battery. Circuits for three (and two) qubits are analysed by considering the generated ergotropy and other measures of performance. We also perform an optimisation of initial system parameters, e.g. initial quantum coherence of one of the qubits affects strongly the efficiency of protocols  and the system's performance as a battery. Finally, we briefly discuss the feasibility for an experimental realization.
\end{abstract}

\maketitle

\section{Introduction}

Advances in technology have enabled
the  construction and manipulation of microscopic systems, opening the door to develop experimental thermal machines which lie far from the thermodynamic limit~\cite{Rosnagel2016,Peterson2019,Klatzow2019,vonLindenfels2019,VanHorne2020,Bouton2021,Lisboa2022,Herrera2023}. Quantum effects and phenomena, such as the entanglement of particles, become extremely important in the description and modelling of these systems. These quantum phenomena can give advantages over classical systems in thermodynamic processes~\cite{Herrera2023,Dillenschneider2009, Rossnagel2014,Vinjanampathy2016,Bera2017}.
Quantum batteries represent an exciting application of quantum principles, with~the potential to revolutionize energy storage for diverse scenarios. This kind of device has been implemented in multiple systems~\cite{Campaioli2017,Ferraro2018,Le2018,Zhang2018,Joshi2022} and ergotropy has been the figure of merit used to quantify its effectiveness~\cite{Yao2022, Binder2015, Campaioli2018}. Quantum batteries may be based on collective quantum phenomena~\cite{Alicki2013,Campaioli2017}. Hence, considering protocols using many-body interactions for these new technologies is of significant importance. Protocols inspired by quantum computing have been proposed for thermal machines, and~the construction and testing of quantum many-body systems supporting the use of quantum logic gates is experimentally realisable using, among~others, trapped ions~\cite{vonLindenfels2019, PhysRevLett.128.110601}, Rydberg atoms~\cite{Wu2021}, and~superconducting qubits~\cite{Abdelhafez2020}. While still in their early stages of development, theoretical and experimental studies have demonstrated the feasibility and potential benefits of quantum batteries~\cite{Alicki2013,Giorgi2015,Binder2015,Perarnau2015,Campaioli2017,Ferraro2018, Shaghaghi_2022}. Understanding their underlying principles and exploring practical implementations is the aim of this work. Replacing full Pauli gate operations with a fractional step-wise protocol has allowed for investigating the presence of quantum friction, i.e.,~the generation of quantum coherence in the instantaneous energy eigenbasis under a non-permuting protocol~\cite{Miller2019,Scandi2020}. The~experimental realization of this protocol has proven that quantum friction induces a violation of the work fluctuation dissipation relation, certifying an additional genuine quantum effect~\cite{Onishenko2024}.

Here, we propose a fractional step-wise protocol with multiple qubits and using experimentally realisable Nth-root
 controlled-not logic gates (NRCGs) for charging and discharging a quantum battery. Repeat applications of NRCGs act as a trotterization of controlled-not logic gates; this has some analogy with collision models used in other works involving the charging of quantum batteries~\cite{Morrone_2023, e23121627, e24060820, PhysRevLett.128.110601}.
The NRCG stepped approach to the dynamics allows access to intermediate ergotropy states typically unobtainable using full controlled-not logic gates. This offers a greater control of the quantity of energy stored in the battery. NRCGs are feasible with current technology, and~we find that, depending on the protocol and initial conditions, the~use of an NRCG may  provide overall more ergotropy than the use of its full controlled-not gate counterpart.
 We investigate the protocol for initial quantum coherences and find that this leads to improved performances as compared to a thermal~state.

\section{Theoretical outline}
We investigate various quantum thermodynamic protocols for charging a quantum battery. We express the scheme in a gate-based approach with quantum circuits of few-qubit systems, initialised and driven by a gate sequence, where we characterize the outcome of the protocol using the amount of ergotropy that has been~generated.

\subsection{System and Quantum~Circuits}
\label{System}
We consider a system composed either by two  (A and B) or by three (A, B, and~C) qubits that interact with each other through the use of NRCGs (Figure~\ref{fig:Circuittwobody}). These gates may generate entanglement between the component qubits. Qubits A and C are initialised each in a thermal state at temperatures $T_{A}$ and $T_{C}$, respectively, where $T_{A} > T_{C}>0$.
The single-qubit thermal state is a Gibbs state defined as~\cite{Lenard1978}
\begin{equation}
 \rho_{Gibbs j} \ = \ (1/Z_{j})\exp^{-\beta_{j} H_{j}},
 \label{eq:gibbsequation}
\end{equation}
where $ Z_{j} = \sum_{i}e^{-\beta_{j} \epsilon_{ij}}$ is the partition function, $\beta_j = (k_BT_j)^{-1}$ is the inverse
temperature parameter, $k_B$ is the Boltzmann constant,  $\epsilon_{ij}$ are the eigenvalues of the single-qubit Hamiltonian. This is given by
\begin{equation}
  H_j \ = \
\begin{pmatrix}
\epsilon_{1j} & 0\\
0 & \epsilon_{2j}
\end{pmatrix},
\label{H_j}
\end{equation}
where $j=A,B,C$. As~reference systems, we consider the ones with each qubit prepared in a thermal state: here, there are no initial quantum coherences.
We compare these with the systems in which qubit B is prepared in a pure state; then, initial quantum coherence will carry through the qubits when a circuit is applied. Qubit B is initialised in a pure state as $\rho_{Pure} = \left|\psi\rangle \langle \psi  \right|$, where $|\psi \rangle = \cos\left(\theta/2\right)|0\rangle + e^{i \phi}\sin\left(\theta/2\right)|1\rangle$ with a value of $\theta$ which ranges from $0 \ to \ \pi$, while $\phi$ ranges from $0$ to $2\pi$. For~this investigation, we will mainly focus on $\phi = 0$ and $\phi = \pi$, though~the full range of $\phi$ values will also be considered. Here, $| 0\rangle$ is the ground state and $| 1 \rangle$ is the excited state. In~this paper, energies are given in units of $\epsilon_{2B}$, which is then set to 1 in all~calculations.

The initial Hamiltonian is non-interacting and of the form,
\begin{equation}
\label{eq:systemhamiltonian}
 H_{System} = \sum_j H_j
\end{equation}
with $H_j$ given by Equation~(\ref{H_j}) and $j=A,B$ or $j=A,B,C$ for two and three qubits, respectively.
We note that Equation~(\ref{eq:systemhamiltonian}) represents the  Hamiltonian for the total system at any time, including at the point of measurement, except~when NRCGs are applied, inducing interactions between~qubits.

\subsection{CNOT and Nth CNOT Root Logic~Gates}\label{CnotGates}
The standard form of the CNOT gate for the two-qubit system is,
\begin{equation}\label{CNOTFORM}
  CNOT_{A,B} \ = \
\begin{pmatrix}
1 & 0 & 0 & 0\\
0 & 1 & 0 & 0\\
0 & 0 & 0 & 1\\
0 & 0 & 1 & 0
\end{pmatrix}, \;\;\;\;\;\;\;\;\;
CNOT_{B,A} \ = \
\begin{pmatrix}
1 & 0 & 0 & 0\\
0 & 0 & 0 & 1\\
0 & 0 & 1 & 0\\
0 & 1 & 0 & 0
\end{pmatrix}.
\end{equation}

where the first subscript represents the control qubit and the second subscript, the target qubit. This type of gate may generate entanglement between two qubits, which adds a level of quantumness to the system. Entanglement of intermediate states, which a CNOT gate may achieve, is known to be necessary to gain a quantum advantage in battery charging operations~\cite{10.1116/5.0184903}. The~NRCG is a method of partially applying a CNOT gate; it is a unitary operation given by~\cite{Nikolov},
\begin{equation}\label{NRCGFORM}
\sqrt[N]{CNOT_{A,B}} \ = \
\begin{pmatrix}
1 & 0 & 0 & 0\\
0 & 1 & 0 & 0\\
0 & 0 &  s & p\\
0 & 0 &  p & s
\end{pmatrix}, \;\;\;\;\;\;
\sqrt[N]{CNOT_{B,A}} \ = \
\begin{pmatrix}
1 & 0 & 0 & 0\\
0 & s & 0 & p\\
0 & 0 &  1 & 0\\
0 & p &  0 & s
\end{pmatrix}.
\end{equation}
Here, $ s = \frac{1}{2} + \frac{1}{2}e^{\frac{i\pi}{N}}$ and $p = \frac{1}{2} - \frac{1}{2}e^{\frac{i\pi}{N}}$.

\begin{figure}[hbt!]
    \includegraphics[width=0.8\linewidth]{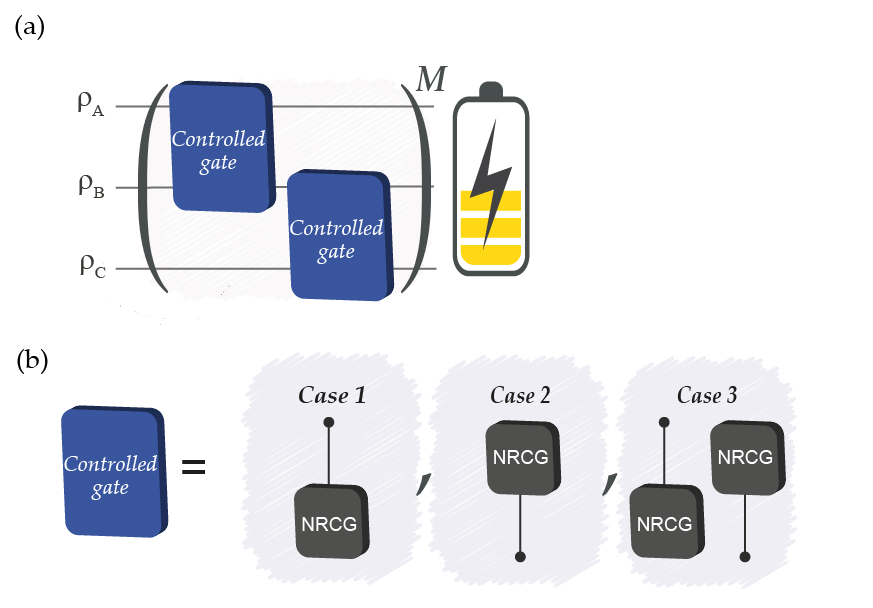}
    \caption{(\textbf{a}) Circuit diagram for the three-qubit protocols: the full protocol includes $M$ iterations, each composed of two blocks of gates (blue squares). If~$M=N$, the~protocol is said to run over a ``cycle''. (\textbf{b}) The protocols considered, i.e.,~case 1, case 2, and~case 3, differ by the set of gates within each `controlled gate' block, as~indicated. 
    A two-qubit protocol emerges from the circuit in panel (\textbf{a}) by removing qubit C and the second controlled gate block.}
    \label{fig:Circuittwobody}
\end{figure}

A cycle is defined as N iterations of the basic circuit, corresponding to $M=N$ in Figure~\ref{fig:Circuittwobody}a.
In the limit of a large $N$, $N$ consecutive applications of an NRCG with the same control and target qubits could be seen as a trotterization of the CNOT gate, aiming at explicitly implementing the gate as an adiabatic dynamic. In~this sense, our protocols give explicit access to intermediate states, e.g.,~allowing for the opportunity to use states with different degrees of entanglement from the end result of the full CNOT~gate.

\subsection{Circuit}
A schematic of the protocols applied is shown in Figure~\ref{fig:Circuittwobody}. We consider systems of two and three qubits and for each of them, we examine three ways of applying interactions, labeled 1, 2, and 3, see Figure~\ref{fig:Circuittwobody}(b).
  Case 1 and 2 examine the dynamics when one of the qubits is never the control qubit, with this being either qubit A or C for the three-qubit systems. Case 3 examines the protocol where all qubits are either a control or target qubit over the course of one iteration. In~all protocols, qubits A and C are always initialised in a thermal state using Equation~(\ref{eq:gibbsequation}), whereas qubit B is initialised in either a pure or a thermal state. This allows for probing how initial quantum coherences affect the system's capabilities as a~battery.

The system is evolved, corresponding to the quantum circuits in Figure~\ref{fig:Circuittwobody} and to the~evolution
\begin{equation}\label{eq:unitary}
   \rho^{(f)}_{sys} = U\rho^{(0)}_{sys}U^{\dagger}
\end{equation}
 where $\rho^{(0)}_{sys}$ is the initial density matrix of the total system  and  $\rho^{(f)}_{sys}$ indicates the total density matrix after one iteration. The~unitary $U$ represents the controlled gates specified in Figure~\ref{fig:Circuittwobody}, as~appropriate to each protocol, with~NRCGs of the form in Equation~(\ref{NRCGFORM}).

We note that in all three-qubit protocols, as~NRCGs are applied to different qubits or alternate different control/target qubits, successive iterations will not generate full CNOT gates. Instead, for~Cases 1 and 2 and the two-qubit system, multiples of $N$ iterations of the circuit will be equivalent to the application of multiple standard CNOT gates; however, fractions of $N$ iterations will allow access to intermediate states in the evolution leading to a~CNOT.

\subsection{Ergotropy}
\label{ergotropy}
Ergotropy is the maximum amount of work that can be extracted from a quantum system by means of a cyclic and  unitary operation~\cite{Allahverdyan2004}. It is the primary measure used to explore the performance of a quantum battery~\cite{Alicki2013}. For~a state $\rho$ of Hamiltonian $H$, the~ergotropy is given by
\begin{equation}
   W_{max} =Tr\left[\rho H\right] - Tr [ \sigma_{\rho} H],
   \label{eq:ergotropy}
\end{equation}
 where  $\sigma_{\rho}$ is the passive state connected to $\rho$ by a unitary transformation  such that
\begin{equation}
    \sigma_{\rho} = \sum_{j} s_{j} | j \rangle \langle j |  ~\mbox{ with }  s_{j+1} \leq s_j,
\end{equation}
 where $\{s_j\}$ are the eigenvalues of $\rho$ and $\{| j \rangle \}$ are the eigenstates of $H$~\cite{Allahverdyan2004}. No work can be extracted from a passive state, and~for all unitaries $U$,
   $Tr\left(\sigma_{{\rho}} H\right) \leq Tr\left( U \sigma_{{\rho}} U^{\dagger} H \right)$~\cite{Allahverdyan2004, Pusz1978}.

\section{Plan of the Paper and Anticipation of Main~Results}
The application of Nth-root gate operation between qubits unlocks a couple of assets, as~we are able to follow the time evolution in a step-wise protocol and investigate the functioning of quantum battery protocols in more detail. Consequently, we consider three different protocols, highlighted in Figure~\ref{fig:Circuittwobody} for varying numbers of iterations and different initializations. For~a fair and meaningful comparison among all these systems and protocols, we will examine four measures of performance for an optimal~battery:
\begin{itemize}
    \item Ergotropy, which describes the maximum amount of work extractable at the end of the protocol;
    \item Ergotropy variation, which is the~difference between the final and initial ergotropies, relevant when the initial state, like that here, is not a passive state;
    \item The ratio between ergotropy and the final energy of the system, which describes the fraction of energy that can be extracted as work;
    \item A figure of merit, which combines the ergotropy variation and the ratio.
\end{itemize}

We find that results are strongly affected by the initial coherence of qubit B, and~the same circuit may lead to minimal or to maximal ergotropy, depending on the setting of the phase. Also, the~overall largest ergotropy variation and figure of merit are obtained from a circuit with high initial coherence.  For~all protocols and all measures of performance, initializing qubit B in a pure state as opposed to a thermal state provides a substantial~advantage.

\section{Results}\label{identical_qubits}

\subsection{Ergotropy Scans for Two- and Three-Qubit~Systems}\label{Ergotropy_scans}
The two-qubit system is initialised with qubit A thermalised in the Gibbs state \mbox{Equation~(\ref{eq:gibbsequation})}. We are using a reservoir where $k_{B}T_{A} = 4\epsilon_{2B}$ . For~the three-qubit system, qubits A and C are initialised  at $k_{B}T_{A} = 4\epsilon_{2B}$ and $ \ k_{B}T_{C} = 0.4\epsilon_{2B}$, respectively. Qubit B is initialised in the pure state  $|\psi\rangle = \cos(\theta/2)|0\rangle + \exp(i\phi)\sin(\theta/2)|1\rangle$.

One cycle is composed by $N$ iterations. We choose $N = 15$, with~this value showing a smooth evolution of the state of the system as iterations are applied. Cycles with $N > 15$ show no improvement but are increasing the run times of the simulations. Cycles with $N < 15$ show increasing discontinuities, moving the evolution closer to the one using a complete CNOT gate ($N=1$). For~details, see Appendix \ref{appendA}, Figure~\ref{fig:convergenceappendix}.

\begin{figure}[hbt!]
    \centering
    \includegraphics[width=15.5cm]{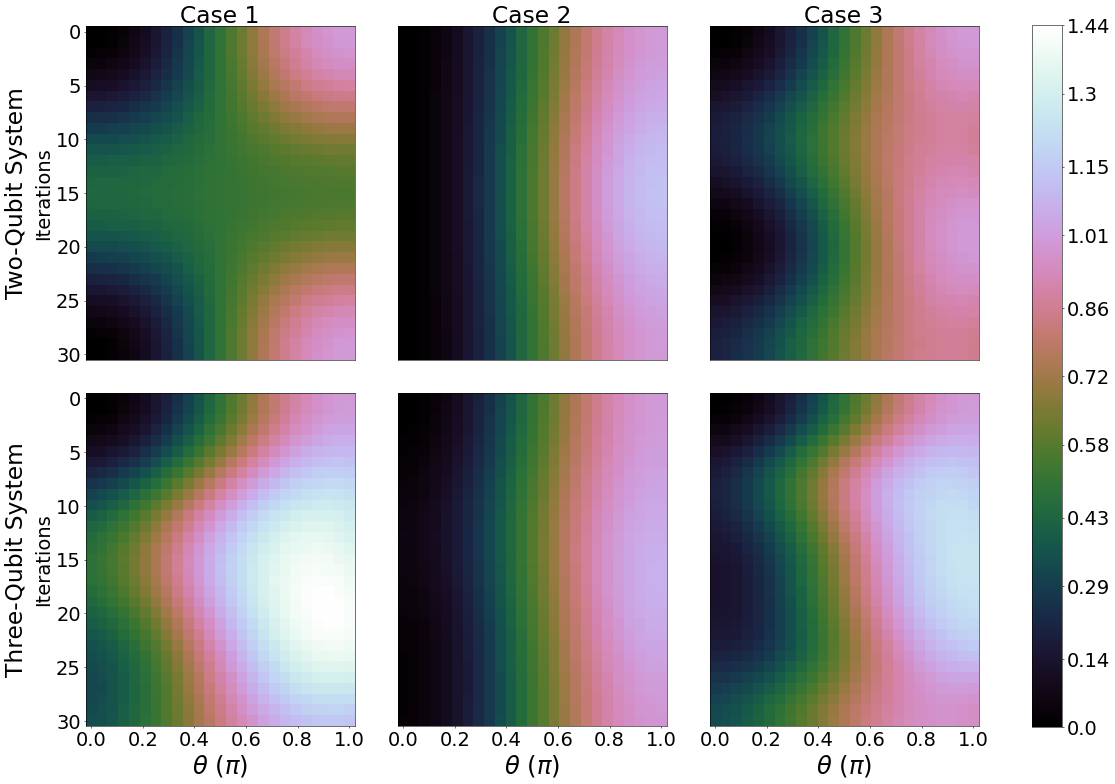}
    \caption{Ergotropy of the total system for $0 \ \leq \ \theta \ \leq \ \pi$ (x-axis) and $0 \ \leq \ $ Iterations $ \ \leq  \ 30 $ (y-axis) for two-qubit systems (first row) and three-qubit systems (second row), where columns from left to right are case 1, 2, and~3, respectively. Brighter shades correspond to a greater value of ergotropy. Parameters are $\epsilon_{1j} = 0\epsilon_{2B}, \ \epsilon_{2j} = 1\epsilon_{2B}$ for $j = A, B$, and~$C$. $k_{B}T_{A} = 4\epsilon_{2B}$  and $k_{B}T_{C} = 0.4\epsilon_{2B}$. $\phi =  \pi$.}
    \label{fig:ErgoScanTotalSystemPhi}
\end{figure}

Each panel of Figure~\ref{fig:ErgoScanTotalSystemPhi} shows results for the ergotropy calculated after 1, 2, $\dots,~2N$ iterations, for~$0\le \theta \le \pi$ and $\phi=\pi$.
Increasing $\theta$ increases qubit B's initial internal energy from  $\epsilon_{1B} = 0$ to $\epsilon_{2B}$.
The first row refers to two-qubit systems, the~second, to results for three-qubit systems, for~all three cases described in Figure~\ref{fig:Circuittwobody}b, as~labelled. We refer to each of the panels as an ``ergotropy~scan''.

The initial ergotropy (corresponding to zero iterations, top line of all panels) mainly depends on the initial energy of qubit B.
The closer $\theta$ is to $\pi$, the larger the initial ergotropy of the total system. This can be seen across all cases for both two- and three-qubit systems.
However, how the battery behaves for an increasing number of iterations depends on both the number of qubits and the circuit chosen, showing charging (ergotropy increasing with iterations) and discharging (ergotropy decreasing with iterations) regimes.
For example, for~case 1, $\theta = \pi$ and iterations increasing from 0 to 15, the~two-qubit system would discharge while the three-qubit system would be~charging.

Cases 1 and 3 show the largest difference in the ergotropy evolution, both with respect to the initial ergotropy value and when comparing the results for the two- and three-qubit systems. It is clear that in cases 1 and 3, there is an optimal number of iterations that would maximise the ergotropy, with~the region of the largest ergotropy reached for three qubits, about 10 to 25 iterations and $\theta\gtrsim 1.9$ radians. For~case 2, we find only a small variation in ergotropy as the system~evolves.

Results with $\phi = 0$ are similar (see Appendix \ref{appendC}, Figure~\ref{fig:ErgoScanTotalSystem}), with~the three-qubit system for both case 1 and 3 showing the most notable variation in ergotropy, though~their maximum ergotropy region gets shifted  towards larger values of $\theta$, $\theta\gtrsim 2.5$ radians, and~they show a lesser variation in $W_{max}$ with~iterations.

\subsection{Comparison Between Circuits with Full CNOT and~NRCGs}\label{CNOT Comparison}

\begin{figure}[hbt!]
    \centering
    \includegraphics[width=15.5cm]{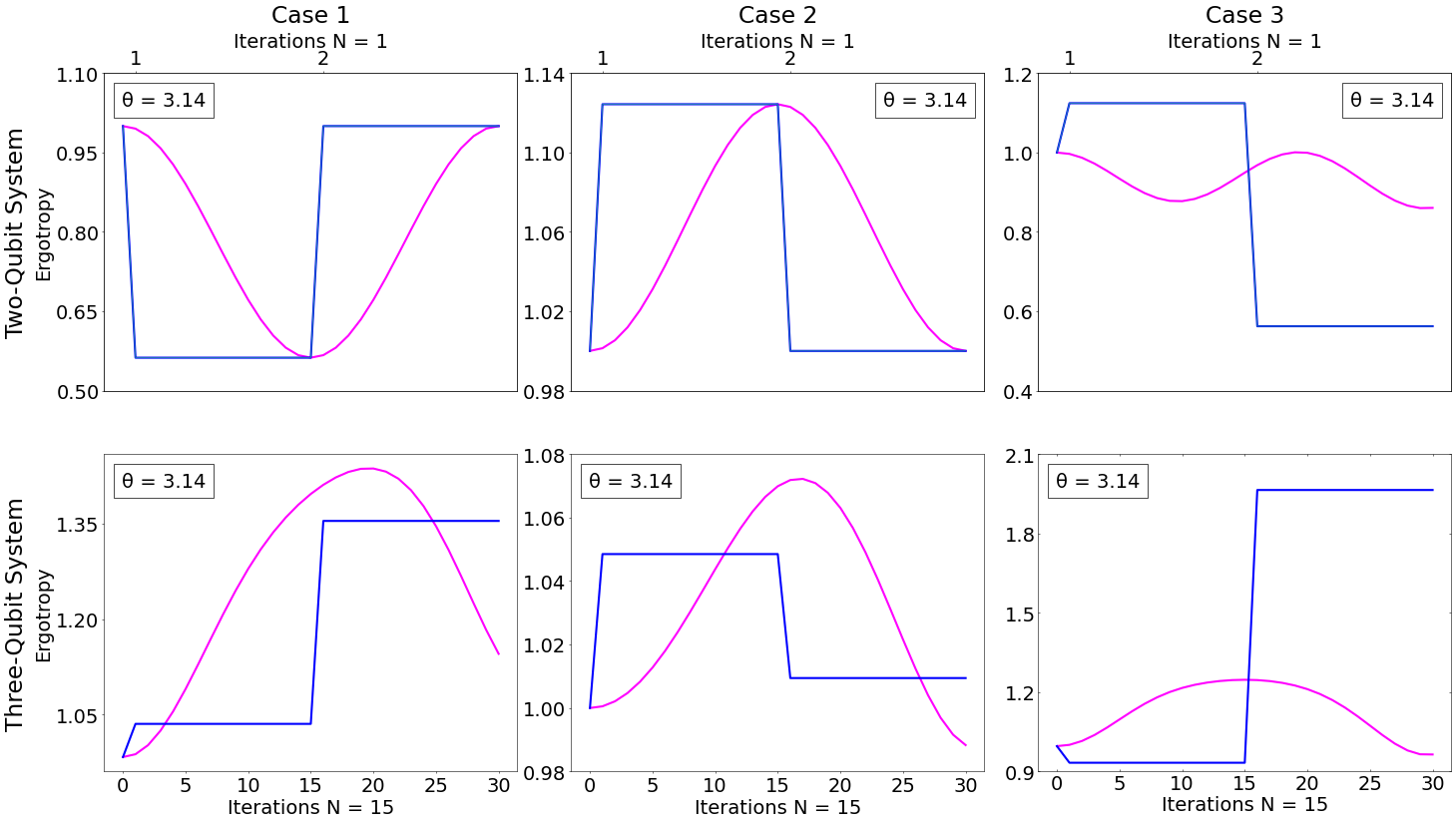}
    \caption{Comparing the ergotropy $W_{max}$ for the NCRG-protocol with $0 \ \leq \ $ N $ \ \leq  \ 30 $ (magenta), and~the  standard CNOT (blue), in~the case of the two-qubit systems (first row) and the three-qubit system (second row), where columns from left to right are cases 1, 2, and~3, respectively. The Y-axis scale differs for each panel. The~chosen value of $\theta$ is indicated in each panel. Parameters: $\Phi = \pi$, $\epsilon_{1j} = 0\epsilon_{2B}, \ \epsilon_{2j} = 1\epsilon_{2B}$ for $j = A, B$, and~$C$; $k_{B}T_{A} = 4\epsilon_{2B}$ and $k_{B}T_{C} = 0.4\epsilon_{2B}$. Magenta line: N = 15; blue line: N = 1. }
    \label{fig:CNOT1COMPPhi}
\end{figure}

NCRGs can be interpreted as a stepped application of a full CNOT gate as explained more thoroughly in Section~\ref{CnotGates}. We then wish to compare whether applying fractions of CNOTs instead of a full CNOT is advantageous in the present case. We compare the ergotropy over two cycles in which iterations have either of one CNOT or N = 15 steps of an NRCG, which is magenta. For~each panel we consider $\theta$ such that the ergotropy of the system considered in the panel reaches its maximum when N = 15. The~top left and top center panels in Figure~\ref{fig:CNOT1COMPPhi}, showing case 1 and 2, for~the two-qubit systems operate with only a single gate per iteration. These particular systems demonstrate well how the NRCGs operate as a slower application of the full CNOT gate, with the systems' ergotropy reaching the same values for both types of gates after one cycle. Circuits with the full CNOT gate allow us to reach only two specific values of ergotropy. However, NRCGs allow access to these and to all intermediate values. Such fine-tuning (increase and decrease) of ergotropy with the system's evolution could prove advantageous when a particular value of ergotropy or the corresponding intermediate state are desired that are in-between the initial and final ergotropy and states generated by a full CNOT~gate.

With the exception of cases 1 and 2 for two-qubit systems, NRCG-based cycles are not equivalent to cycles from corresponding circuits with CNOTs, even when considering the end results, see Figure~\ref{fig:CNOT1COMPPhi}. Apart from the easy reach of intermediate values of ergotropy, we can see two other ways in which the use of NRCGs could be advantageous: (i) at points where the ergotropy of the total system has a larger value than that of the standard CNOT gate within the same cycle; or (ii) when the maximum ergotropy reached using NRCGs is higher that the one of the corresponding CNOT gate circuits. The~first possibility can be observed in all panels of Figure~\ref{fig:CNOT1COMPPhi}; the second can be seen in the three-qubit cases 1 and~2.

Similar results are found for the case of qubit B, initialised in $\Phi = 0$. We note that a larger value of the ergotropy may be reached after the two cycles considered in Figure~\ref{fig:CNOT1COMPPhi}.
 
\subsection{Ergotropy Variation for  Two- and Three-Qubit~Systems}\label{Ergotropy_Difference}

Our systems do not start from a passive state, so the ergotropy $W_{max}$ at the end of the process may be influenced by the initial one. This is not a simple relation, as~the systems studied undergo `discharging' and `charging' processes as the number of iterations increase, as~discussed~below.
\begin{figure}[hbt!]
    \centering
    \includegraphics[width=15.5cm]{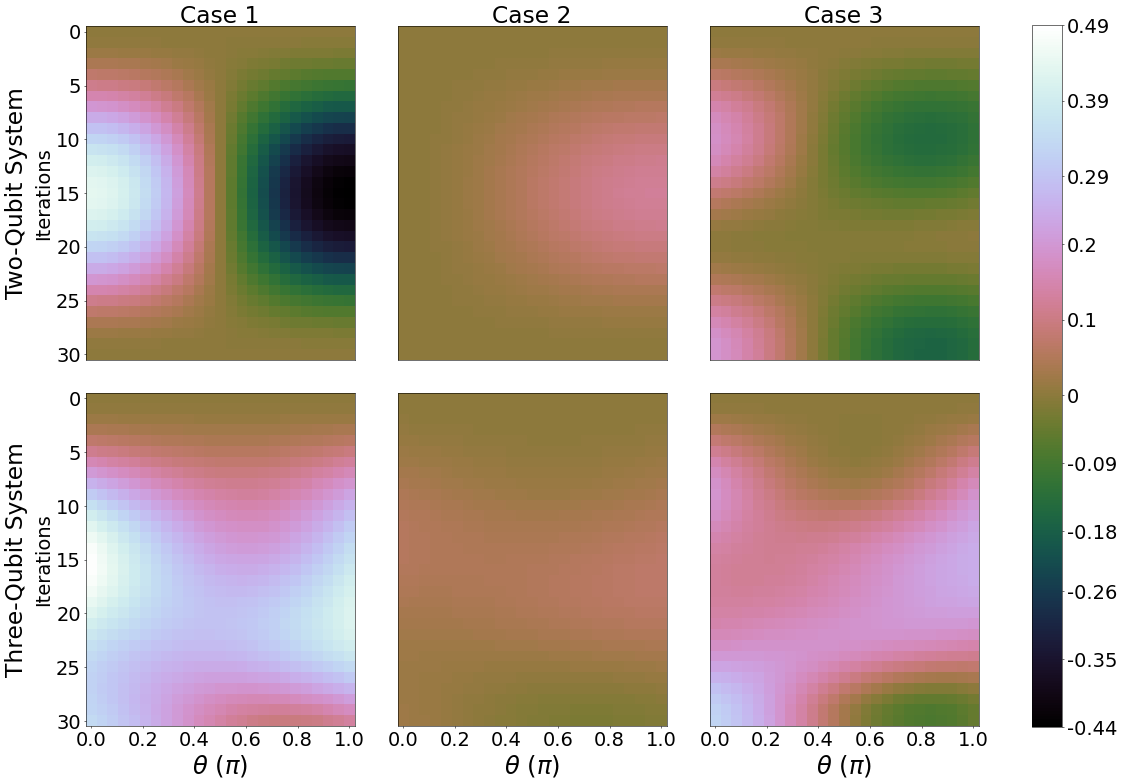}
    \caption{{Ergotropy}
 variation $\Delta W_{max}$ for $\phi = 0$. Otherwise, we have the same parameters as in Figure~\ref{fig:ErgoScanTotalSystemPhi}.}
    \label{fig:DeltaErgTot}
\end{figure}
\unskip

\begin{figure}[hbt!]
    \centering
    \includegraphics[width=15.5cm]{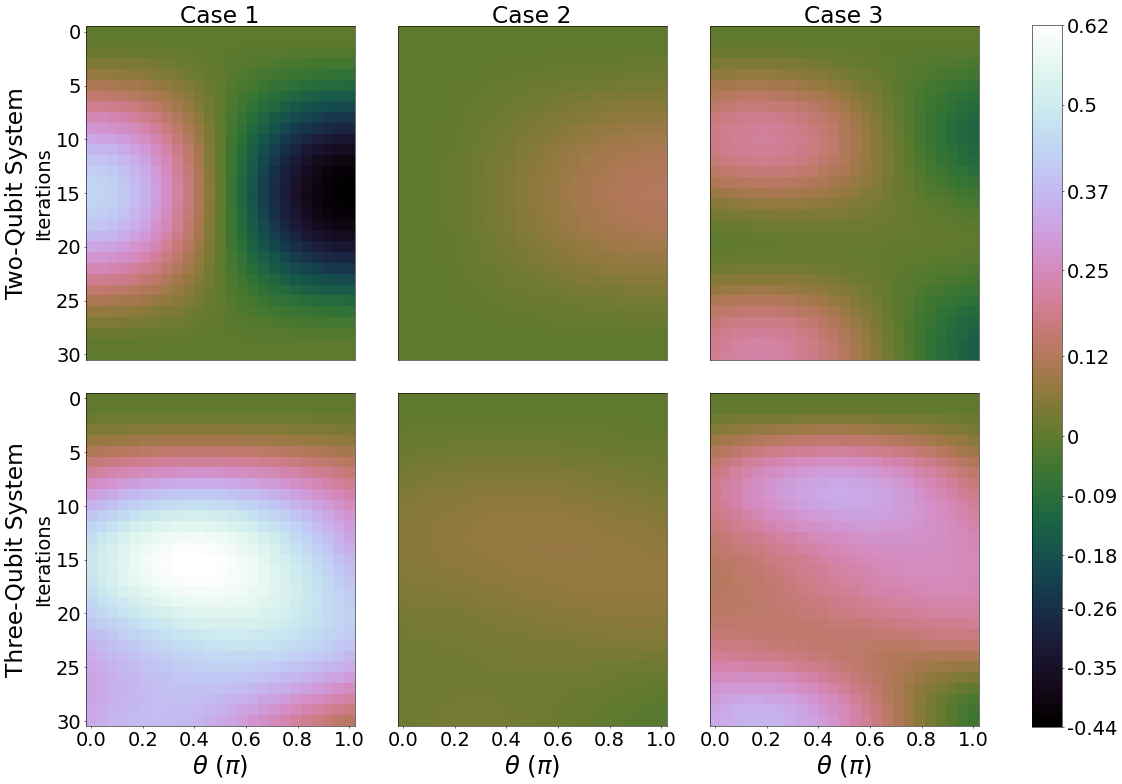}
    \caption{{Ergotropy}
 variation $\Delta W_{max}$ for $\phi = \pi$, with the same parameters as in Figure~\ref{fig:ErgoScanTotalSystemPhi}.}
    \label{fig:DeltaErgTotPhi}
\end{figure}
In addition, when considering $W_{max}$ (Figure~\ref{fig:ErgoScanTotalSystemPhi}), a~direct comparison between the two- and three-qubit systems is not easy to perform: when all common parameters are the same, the~three-qubit system inherently starts from more energy and ergotropy than the two-qubit system by merit of having more qubits. For~the dynamics considered, this almost always translates into acquiring more ergotropy by the end of the process (compare the first and second row in Figure~\ref{fig:ErgoScanTotalSystemPhi}). From~this, we can argue that looking at the maximum ergotropy may not be a fair comparison between systems with different quantities of qubits. For~these reasons, we introduce the ergotropy variation ($\Delta W_{max}$) which is defined by the following~equation:
\begin{equation}\label{eq:ergodifference}
    \Delta W_{max} = W_{max}^{(f)} - W_{max}^{(0)},
\end{equation}
where $W_{max}^{(f)}$ is the ergotropy of the final state and $W_{max}^{(0)}$ is the ergotropy of the initial state of the total system. Equation~\eqref{eq:ergodifference} allows the identification of two regimes with respect to the initial state, charging or discharging. A~charging regime is characterised by a positive value of ergotropy variation and a discharging regime is characterised by a negative value of ergotropy variation. $\Delta W_{max}$ allows for a clearer comparison across different systems. For~example, an equal or greater positive change in the two-qubit system compared to the corresponding three-qubit system may suggest that an easier-to-implement two-qubit circuit may be more suited to accrue energy as a~battery.

$\Delta W_{max}$ is evaluated and plotted for $\phi = 0$, see Figure~\ref{fig:DeltaErgTot} and $\phi = \pi$, see Figure~\ref{fig:DeltaErgTotPhi}. The~corresponding panels for three-qubit systems show a dramatically different landscape, with~the maxima no longer being constrained to $\theta \approx \pi$. We can see for case 2 that for the two- and three-qubit system, $\Delta W_{max}$ is small when compared to case 1 and 3, showing that it is not the optimal case for increasing or decreasing ergotropy from the initial state of the system. Cases 1 and 3 in Figure~\ref{fig:DeltaErgTot} display much larger peaks and troughs for $\Delta W_{max}$ with respect to case 2, indicating a stronger performance as a quantum battery with well-defined charging and discharging regimes. Comparison of cases 1 and 2, which correspond to inverted gates, indicates that having qubit A (thermalised initially at the higher temperature) be a control qubit produces more favourable changes in ergotropy. Also, the~introduction of an extra two gates per iteration in case 3 fails to provide higher maxima in $\Delta W_{max}$ than case 1. Across all panels in Figure~\ref{fig:DeltaErgTot}, the charging  regions (positive $\Delta W_{max}$) are dominating over the discharging region, which is of advantage when looking to optimal conditions for operating a quantum battery.
When qubit B is initialised with $\phi = \pi$ for three-qubit systems and cases 1 and 3, the maximum variation of ergotropy changes from 0.49$\epsilon_{2B}$ to 0.62$\epsilon_{2B}$, which is a 26.5\% improvement in performance. For~most cases, the~maximum $\Delta W_{max}$ regions now lie closer to $\theta = \pi/2$. This shows that the extra initial coherence enhances the charging performance in most circuits considered, and~especially so for case 1 and 3 and three-qubit~systems.

The ergotropy variation $\Delta W_{max}$ should not be the only quantity to consider when looking at the operating parameters for a quantum battery. A~fair additional question could be how much of the final systems' energy can be extracted as work.  We will look at this, the~'ergotropy ratio', in~the next~section.

\subsection{Ergotropy Ratio for Two- and Three-Qubit~Circuits}\label{Ergotropy Ratio}

Here, we introduce the ergotropy ratio which is defined as,
\begin{equation}\label{eq:ErgRat}
    W_{ratio} = \frac{W_{max}}{Tr[\rho^{(f)} H]},
\end{equation}
where $W_{max}$ is the ergotropy (Equation (\ref{eq:ergotropy})) and $Tr[\rho^{(f)} H]$ is the internal energy of the system in its final state. The~ergotropy ratio takes a value between 0 and 1, where a value closer to 1 signifies a larger quantity of extractable energy. The~ergotropy ratio is a good measure of how efficiently a battery is performing~\cite{PhysRevLett.122.047702}. With~this in mind, here, $W_{ratio}$ is introduced to look for optimal initial conditions for an efficient quantum~battery.

\begin{figure}[hbt!]
    \centering
    \includegraphics[width=15.5cm]{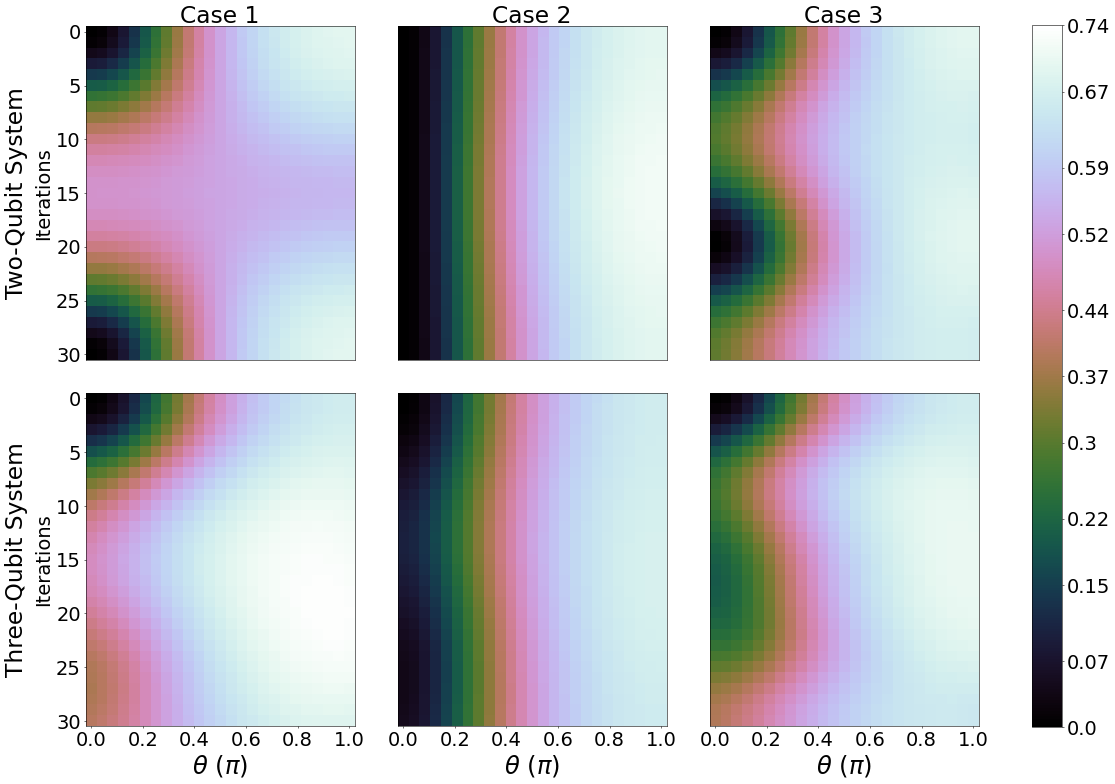}
    \caption{Ergotropy ratio $W_{ratio}$ for $\Phi = \pi$, with the same parameters as in Figure~\ref{fig:ErgoScanTotalSystemPhi}. }
    \label{fig:ERGRATPhi}
\end{figure}

Figure~\ref{fig:ERGRATPhi} shows the ergotropy ratio for the total system for the same parameters as those in Figure~\ref{fig:ErgoScanTotalSystemPhi}, in~particular, qubit B is initialised with $\Phi = \pi$. For~all cases, we obtain a greater value of ergotropy ratio when qubit B is initialised with a large value of $\theta$, but~not necessarily $\theta=\pi$. The~maximum values for $W_{ratio}$ across all panels in Figure~\ref{fig:ERGRATPhi} are similar, suggesting no particular advantage with using a three- instead of a two-qubit system. When initialising qubit B with $\Phi = \pi$ (see Figure~\ref{fig:ERGRATPhi}), we observe a marked increased in areas of high $W_{ratio}$ for values of $\theta$ closer to $\pi/2$ for both two- (case 3) and three- (cases 1 and 3) qubit systems, supporting the advantage from a larger initial coherence. For~comparison, results with qubit B initialised at $\Phi = 0$ are shown in Figure~\ref{fig:ERGRAT}, Appendix~\ref{appendC}.

While $W_{ratio}$ is a good indicator for which initial conditions lead to a larger percentage of total energy that can be extracted as work, it does not show if that large percentage also  corresponds to a region of high charging: indeed, $W_{ratio}$ and $\Delta W_{max}$ have quite different behaviours (e.g., compare  Figures~\ref{fig:DeltaErgTot} and \ref{fig:ERGRAT}). Ideally, one would like to optimize both quantities at the same time to quantitatively assess how well the initial conditions, system, and~circuit considered perform as a battery. In~the next section, we will introduce a related figure of~merit.

\subsection{Figure of Merit for Optimal Battery, Two- and Three-Qubit~Systems}\label{figure_of_merit}

The figure of merit we propose to find the optimal circuit, number of iterations, and initial conditions for charging a quantum battery has the form
\begin{equation}\label{eq:FigureOfMerit}
    FoM (\theta, \Phi, \mbox{iter}) = \Delta W \times W_{ratio}.
\end{equation}

This combination of quantities shows where the best balance of energy that can be extracted (ergotropy ratio) and the ergotropy increase from the initial condition (ergotropy difference) lie, giving a clearer indication of the optimal~system.

\begin{figure}[hbt!]
    \centering
    \includegraphics[width=15.5cm]{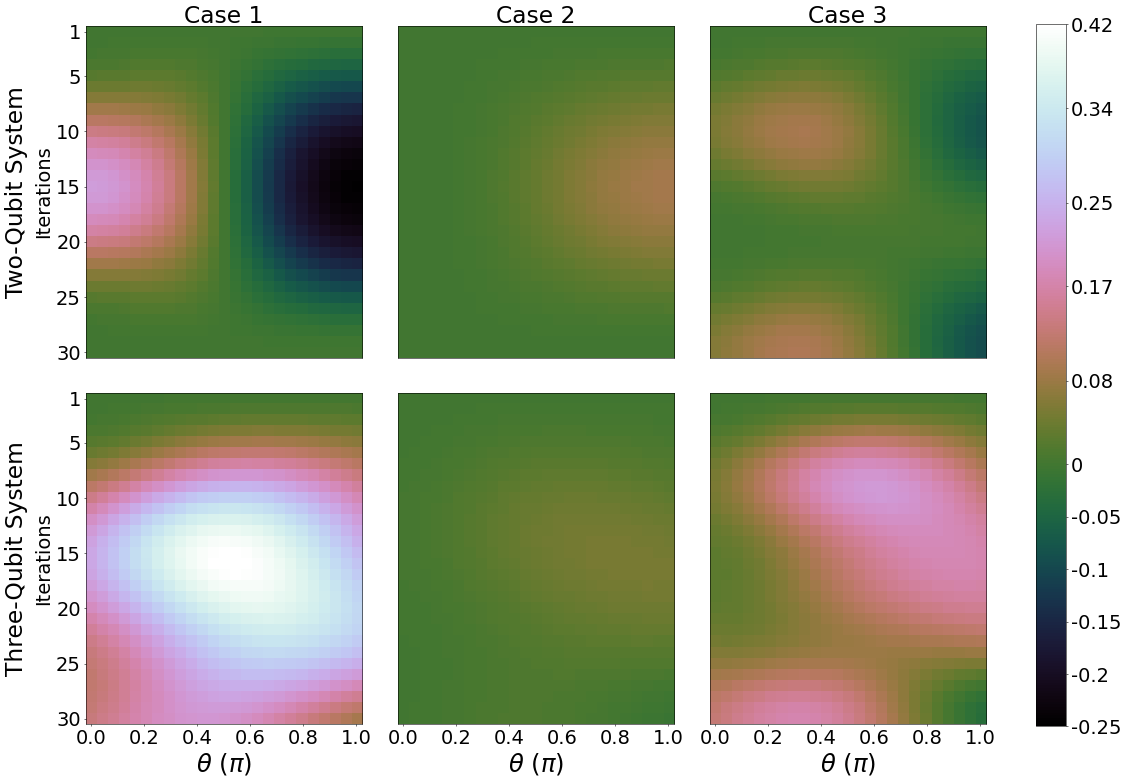}
    \caption{ {Figure}
 of merit, FoM, with~$\Phi = \pi$, with the same parameters as Figure~\ref{fig:ErgoScanTotalSystemPhi}.}
    \label{fig:FiguresOfMeritPhi}
\end{figure}

In Figure~\ref{fig:FiguresOfMeritPhi}, we plot $FoM$ for the same parameter scans of Figures~\ref{fig:DeltaErgTotPhi} and \ref{fig:ERGRATPhi}. In~particular, qubit B is initialised with $\Phi=\pi$.  The~results corresponding to $\Phi=0$ are plotted in Appendix~\ref{appendC}, Figure~\ref{fig:FiguresOfMerit}. The~topography of these graphs  tends to mirror that of the variation in ergotropy, which is in~general more variable with iterations than that of the ergotropy ratio. Interestingly, when $\Phi = 0$ (Figure~\ref{fig:FiguresOfMerit}), we see the highest FoM value regions when $\theta\approx 0$ and $\theta\approx\pi$, suggesting that the system performs better as a battery when qubit B is initialised closer to a state with no coherences ($\Phi = 0$, with~$\theta = 0$ or $\pi$). When comparing this to Figure~\ref{fig:FiguresOfMeritPhi} ($\Phi = \pi$), the opposite is true, with~the greatest value of FoM occurring when $\theta = \pi /2 $. The~maximum value for the FoM is also 35\% greater when initialising qubit B with $\phi = \pi$ with the system demonstrating a better performance when introducing initial~coherences.

\begin{figure}[hbt!]
    \centering
    \includegraphics[width=15.5cm]{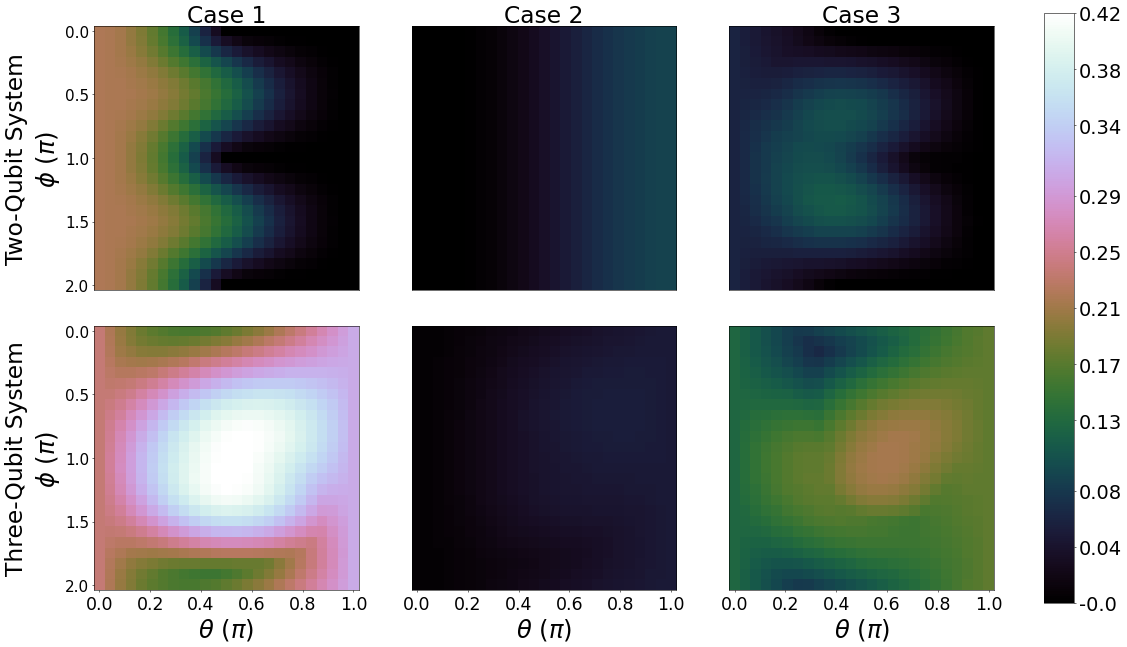}
    \caption{ {First}
 row: Figure of merit of the total system in units of $\epsilon_{2B}$ for $0 \ \leq \ \theta \ \leq \ \pi$ (x-axis) and $0 \ \leq \ $ $\phi$ $ \ \leq  \ 2\pi $ (y-axis) for two-qubit systems where columns from left to right are case 1, 2, and~3, respectively. Parameters: $\epsilon_{1j} = 0\epsilon_{2B}, \ \epsilon_{2j} = 1\epsilon_{2B}$ for $j = A, B$, and~$C$. $k_{B}T_{A} = 4\epsilon_{2B}$ and $k_{B}T_{C} = 0.4\epsilon_{2B}$. Brighter shades correspond to a greater value of the figure of merit. Second row: The same parameters as the first row but for three-qubit~systems. }
    \label{fig:FiguresOfMeritPT}
\end{figure}

We now look at the dependence of $FoM$ on both $\theta$ and $\phi$, with~$0\le\phi<2\pi$ and $0\le\theta<\pi$. The~values plotted in Figure~\ref{fig:FiguresOfMeritPT} are the overall maximum $FoM$ over 30 iterations for each given combination of $\theta$ and $\phi$.
Case 1 with the three-qubit system displays a clear advantage over the others (bottom left panel). Here, there is a large parameter region with a maximum value of $FoM\approx 0.42$, which is approximately double the next best cases (case 1 with two-qubits and case 3 with three-qubits) and which lies in the range of $ 0.63 \ \lesssim \ \theta \ \lesssim \ 2.51$ and $ 1.57 \ \lesssim \ \phi \ \lesssim \ 4.71$.  When looking at the two-qubit systems for case 1 and case 3, we see that the value for the FoM has a double-peak pattern centered where the initial $\phi$ is either $\pi/2$ or $3\pi/4$ and dropping to 0 for large values of $\theta$. For~two qubits, FoM acquires the highest values in case 1, for~$\theta \approx 0$, so with negligible initial coherence.
For the three-qubit system, cases 3 and 1 have a similar trend, with~the maximum value region around $\theta = \pi/2$ and $\phi = \pi$. This is a state with large initial coherence. Our results show for these cases a strong dependence of the FoM maximum value on $\phi$, which is observed to be as important as $\theta$ when selecting optimal initial conditions.
The least favourable circuit corresponds to case 2 (column 2): while it may present among the highest ergotropy ratios (Figures~\ref{fig:ERGRAT} and~\ref{fig:ERGRATPhi}), it shows an ergotropy variation close to zero (Figures~\ref{fig:DeltaErgTot} and~\ref{fig:DeltaErgTotPhi}) which translates into a poor figure of merit. Interestingly, in~contrast to the other cases, case 2 may perform better as a two-qubit system rather than a three-qubit~system.

\section{Comparison with Overall Thermal~Initialization}\label{thermal-purecomparison}

 In previous sections, we have  seen that initial coherences tend to favor higher performances. Here, we compare the systems analysed so far
 with systems in which qubit B is initialised in  a thermal state. Our results show a clear advantage of having qubit B initialised in a pure~state.

\begin{figure}[hbt!]
    \centering
    \includegraphics[width=15.5cm]{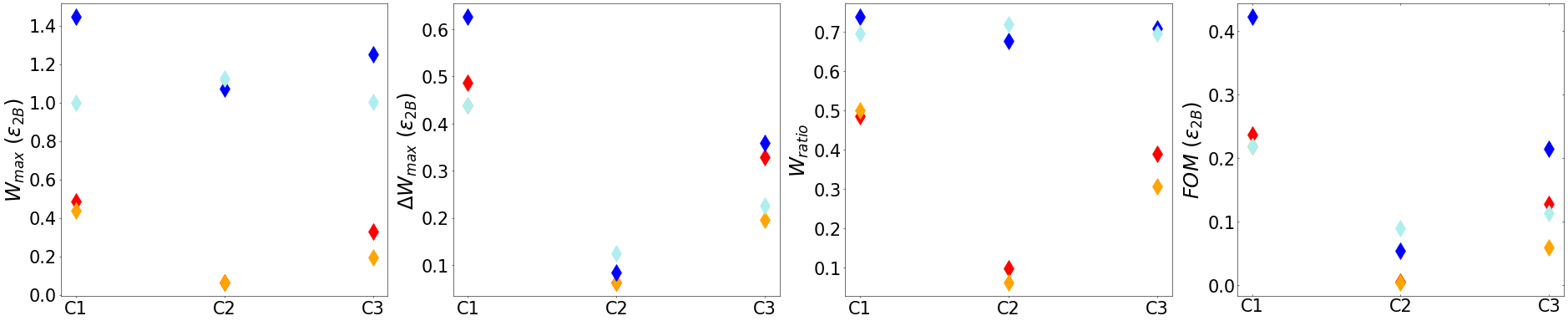}
    \caption{Comparison of performance measures: maximum ergotropy $W_{max}$ (first panel), maximum variation in ergotropy $\Delta W_{max}$ (second panel), maximum ergotropy ratio $W_{ratio}$ (third panel), and maximum figure of merit FoM (fourth panel) of the total system for case 1 (C1), case 2 (C2), and case 3 (C3), as~labelled.  Three-qubit system (blue diamonds), two-qubit system (light-blue), Three-qubit system with B$_{thermal}$ (red) , two-qubit system with B$_{thermal}$ (orange). Parameters: $k_{B}T_{A} = 4\epsilon_{2B}$ and $k_{B}T_{C} = 0.4\epsilon_{2B}$. Maxima calculated over two cycles with $N=15$,  $0\le\phi\le 2\pi$ and $0\le\theta\le\pi$.  }
    \label{fig:MainComps}
\end{figure}

 In Figure~\ref{fig:MainComps}, dark and light blue diamonds correspond to the original systems, i.e.,~ with B initialised in a pure state, for~three- and two-qubit systems, respectively, while red and orange diamonds map the systems with B initially being thermal (B$_{thermal}$  systems) for~three- and two-qubit systems, respectively. The~figure shows the results for the optimal initial configurations producing ergotropy (first panel), maximum ergotropy generation (second panel), the maximum ergotropy ratio (third panel), and~maximum FoM (fourth panel) for case 1, case 2, and case 3, as~labelled. Qubits $A$ and $C$ are always prepared in the thermal state Equation~(\ref{eq:gibbsequation})  with fixed temperatures $k_{B}T_{A} = 4\epsilon_{2B}$ and $k_{B}T_{C} = 0.4\epsilon_{2B}$ (only $k_{B}T_{A}$ applies when dealing with a two-qubit system). We note that each point in a panel may correspond to a different initial condition, depending on which initial value of $\theta$ and $\Phi$ will give the overall maximum of that particular quantity over 30~iterations.

  To find the optimal values for the original systems, parameter scans for qubit B are performed for $0\le\phi\le 2\pi$ and $0\le\theta\le\pi$. For~the initialisation of the B$_{thermal}$ systems, we scan the initial state of qubit B between the zero-temperature state (ground state) and the maximum energy and entropy state
 $ \rho_B \ = \
\begin{pmatrix}
1/2 & 0\\
0 & 1/2
\end{pmatrix}$.
 We vary the temperature in steps between the lowest and highest energy states, with~the generic temperature defined by $k_B T_B = \langle E_B \rangle = Tr\left[\rho_B H_{0}\right]$.

The systematic advantage we observe in the original systems (blue and light blue diamonds) comes from the quantum coherences in the initial state of qubit B, as~can be seen, e.g.,~from Figures~\ref{fig:FiguresOfMeritPhi} and~\ref{fig:FiguresOfMeritPT}. This quantum advantage translates into a maximum ergotropy (first panel) about an order of magnitude higher for case 2, and~two to five times higher for cases 1 and 3~.

For the original systems, three-qubit set-ups outperform two-qubit ones for cases 1 and 3, while the opposite occurs for case 2.
Case 1 for the original system with three qubits demonstrates the best performance across all~measures.

\section{Experimental~Feasibility}
The protocol requires several ingredients, each of which will be discussed separately: it is not challenging for any existing trapped ion quantum processor to operate a small set of 2 or 3 qubits. The~initialization of qubit B into a superposition state is also a simple task for any quantum processor. The~inizialization to Gibbs states of qubits A and C, however, is a non-standard operation, but~it has been realized in some trapped ion quantum setups~\cite{vonLindenfels2019,Onishenko2024,Stahl2024}. The~step-wise execution of the gate, namely the Nth root two-qubit gate, requires the accurate calibration of the qubit diving, which is possibly a challenge when N is large. However, a~step-wise execution appears to be feasible in trapped ion quantum processing, since recently it has been realized but for a single qubit operation~\cite{Onishenko2024} with N up to 10, and~it is also well established for trotterized simulations like, for example, in the trapped ion quantum simulation of a high-energy system~\cite{Martinez2016}. Finally, for~the characterization of the quantum battery output, all methods for state tomography are readily available in trapped ions but also in any of the other platforms for quantum~processors.

\section{Discussion}

A benefit of this current investigation is that NRCGs and the two--three qubit circuits we propose could be realised experimentally on a number of different platforms by using computational gates;  this  is an important next step in the development of quantum batteries, as noted by Campaioli et al.~\cite{Campaioli2018}. When looking at other investigations that inspired this work, the focus is generally on maximum work extraction, which we first looked at in Section~\ref{Ergotropy_scans}. The~importance of initial coherences for the ergotropy was noted Refs.~\cite{PhysRevLett.125.180603, PhysRevE.105.014101}. While we find that the ergotropy is positively influenced by initial coherence, the~other measures of performance we consider are even more affected, as~summarized below. The~power of charging, and by extension the efficiency~\cite{Binder2015,Tacchino2020,Julia,Campaioli2017}, are also used in quantum battery research. The~power is investigated in Appendix \ref{appendB}.

Our investigation compares a set of three different protocols with two or three identical qubits as working fluid. One of the qubits is initialised in a pure state. The idea to~establish which protocol and system is best as a battery under which circumstances has led us to consider a set of measures of performance. Hence, we focus not only on how much work or power can be extracted, but~also on which fraction of the system's internal energy is extractable and characterised this by the ergotropy ratio (Equation (\ref{eq:ErgRat})), used also in~\cite{PhysRevLett.122.047702}.  We find that regions with better performance depend on the initialisation of the system just as much as the method used for charging, generally favouring initial qubit B states where $\theta > \pi/2$ and hence with some degree of population inversion.
However, comparison with results from corresponding initial states where the off-diagonal elements of qubit B are set to $0$  demonstrates that initial coherences significantly enhance the ergotropy ratio for $\theta < \pi/2$. 

 The difference between initial and final ergotropies $\Delta W$ is introduced as a performance indicator. When looking at optimising initial conditions for battery charging, this is arguably a  metric that is more important than just the total ergotropy, especially when initialising the system in a state other than the ground state. With~this, we can discriminate clear winners among the cases implemented, with~case 2 under-performing against cases 1 and 3. This indicator also clarifies charging and discharging regimes, with positive regions corresponding to charging and negative to discharging. We find that initial coherence in qubit B is definitely beneficial to charging, improving the maximum ergotropy difference by 26.5\% (compare results for $\phi = 0$ and $\phi = \pi$).
Results for the power in Appendix \ref{appendB} confirm the importance of initial coherences for enhanced~performances.

With the aim of identifying systems that would be good overall performers, we then introduce a figure of merit, FoM, combining the ergotropy ratio and the variation in ergotropy. By~scanning all initial conditions, the~FoM allows us to identify the protocol of case 1 with three-qubits as the overall best for use as a battery. The~FoM of this system has a marked dependence on the initial condition, favouring  initialisations with the largest initial coherence in qubit B, providing us with a clear quantum~advantage.

In the last part of this paper, we  compare performances for all protocols with corresponding systems with qubit B initialised in a thermal state. The~main takeaway from these comparisons is that introducing a pure-state initial component is advantageous across the board, and~for all measures of performance, see Figure~\ref{fig:MainComps}. Initial coherences lead to overall maxima for some measures of~performance.

Further investigations could expand this work by
exploring the effect of intermediate measurements on the protocols' efficiency~\cite{Son2022} and
utilizing alternative Nth root computational gates and different quantum thermal machines. We found that, depending on the protocol, the~best performer could be either the two- or the three-qubit system: future research could explore the scaling (and its consistency) for increasing the number of qubits.
Another avenue of future investigation could be the use of a collective bath~\cite{PhysRevLett.132.210402} and of global collective charging~\cite{PhysRevLett.128.140501}  to verify if, in~our case, they would improve~performances.

A challenge that this system and others could face experimentally is the loss of coherences, whether from dissipation
 or through direct measurement~\cite{Son2022}.  How ergotropy and related measures of performance would be affected would depend on the dissipation mechanisms specific to the hardware. However, as~hardware for quantum computers has already been optimized to perform a very large number of gates within the relevant decoherence times, quantum batteries based on quantum circuits should also share this~advantage.

\section{Conclusions}
In this paper we have shown that circuits containing Nth-root CNOT gates could be advantageously used as quantum batteries.
We evaluate the circuits' performance using five different measures---ergotropy, power, ergotropy difference and ratio, and~a figure of merit---and considering two- and three-qubit circuits. Our results show consistently that having one of the qubits initialised in a pure state is highly advantageous with respect to using a thermal state with the same energy: high performances are strongly and positively influenced by initial quantum coherences.
For any given initial condition, and~at difference with standard CNOT gates, iterating circuits with Nth-root CNOT gates allows access to a fine distribution of values of ergotropy: this flexibility could be advantageous for tailoring the same battery to different working needs.
The systems and protocol we propose are experimentally feasible with the current~technology.
\vspace{6pt} 

\noindent Conceptualization, E.F., F.S. and I.D.; methodology, E.F. and I.D; software, E.F.; validation, E.F. and I.D.; formal analysis, E.F., M.H. and I.D.; investigation, E.F. and I.D.; resources, I.D.; data curation, E.F.; writing---original draft preparation, E.F. and I.D.; writing---review and editing, E.F., M.H., F.S. and I.D.; visualization, E.F., M.H.; supervision, I.D.; project administration, I.D. All authors have read and agreed to the published version of the manuscript.

\noindent EJF acknowledges support from EPSRC,  grant number is EP/T518025/1; FSK acknowledges funding by the DFG within FOR 2724
 
\noindent The authors declare no conflicts of interest.

\appendix
\section[\appendixname~\thesection]{}\label{appendA}
\begin{figure}[hbt!]
    \centering
    \includegraphics[width=15.5cm]{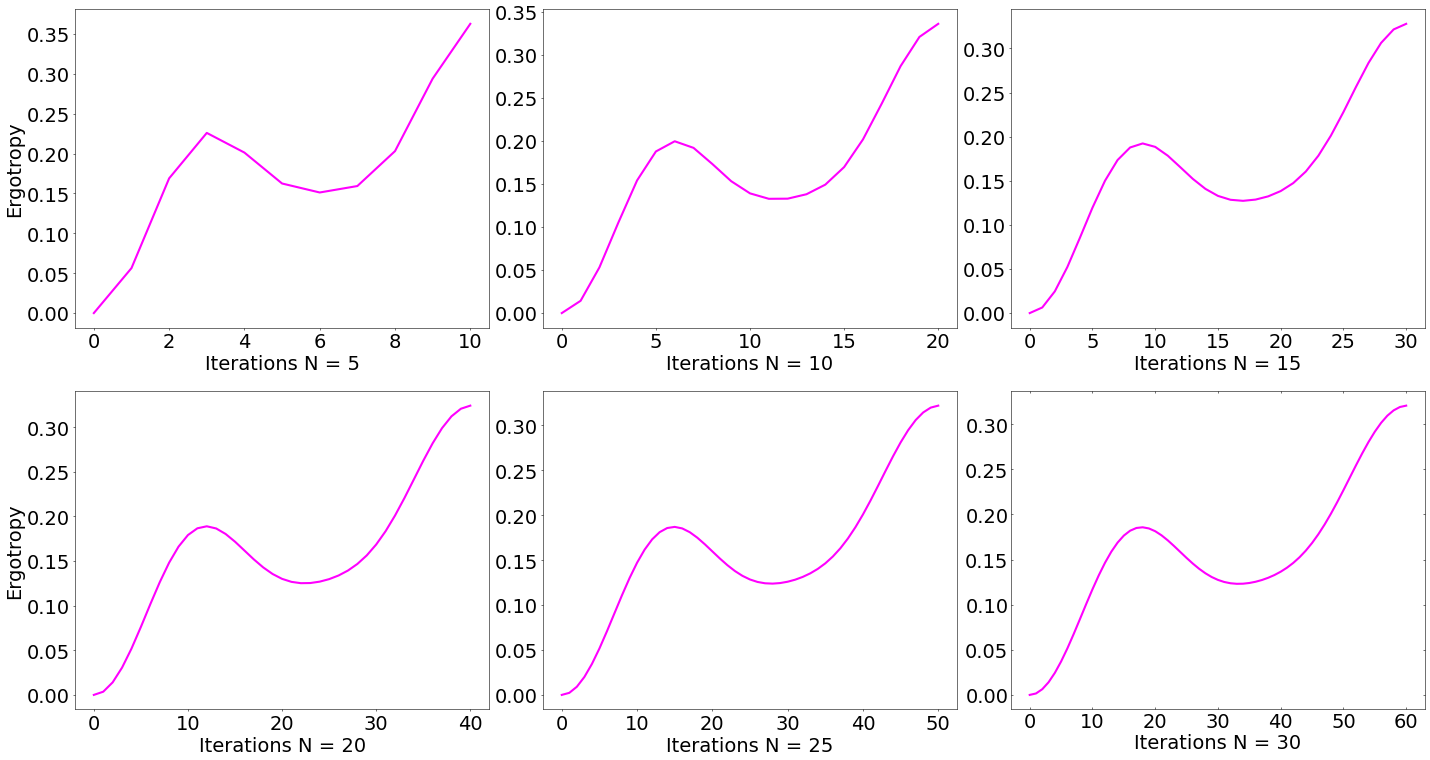}
    \caption{ Ergotropy $W_{max}$ of the total system for $0 \ \leq \ $ iterations $ \ \leq  \ 2N $. For~each panel, $N$ is indicated on the x-axis; protocol corresponding to  case 3 with three-qubit systems. Note that each panel corresponds to two cycles of iterations. Parameters: $\epsilon_{1j} = 0, \ \epsilon_{2j} = 1$ for $j = A, B$, and~$C$.}
    \label{fig:convergenceappendix}
\end{figure}

Figure~\ref{fig:convergenceappendix} shows how changing $N$ affects the ergotropy for case 3 and the three-qubit system. We observe a ``smoothing'' of the evolution of the ergotropy for increasing $N$.

\section[\appendixname~\thesubsection]{}\label{appendB}

Adapting from Refs.~\cite{Binder2015,Campaioli2017}, the~power of charging for our protocols up to the $\alpha$ iteration   is
\begin{equation}\label{eq:powerwork}
    P_{\alpha} = \frac{\langle W_{\alpha} \rangle}{T_{\alpha}},
\end{equation}
where $T_{\alpha}= \alpha / M$ is the process duration up to the $\alpha$ iteration and $M$ is the total number of iterations considered (here, $M=30$ corresponding to two cycles). For~$\alpha = M$, $T_{\alpha} = 1$. $\langle W_{\alpha} \rangle$ is the average work defined as $\langle W_{\alpha} \rangle = E^{\alpha} - E^{i}$, where $E^{\alpha} = Tr \left[ \rho^{\alpha} H \right]$.

When evaluating the power of a quantum battery, there is more depth than just the difference in the average internal energy of the system. While the average internal energy plays a role, this is not a one-to-one comparison with the ergotropy, which as thoroughly discussed in the main article, being a better measure for quantum battery performance. With~that in mind, we introduce,
\begin{equation}\label{eq:powerergdif}
    P_{\Delta W_{max}} = \frac{\Delta W_{max}}{T_{\alpha}},
\end{equation}
where $P_{\Delta W_{max}}$ is the power of charging or discharging and $\Delta W_{max}$ is the variation in ergotropy up to the $\alpha$ iteration. Equation~(\ref{eq:powerergdif}) characterises the rate at which the system generates ergotropy. The~total, ergotropy is not used as we are interested in charging from some initial state ($\rho^{i}$) that has a non-zero~ergotropy.

\begin{figure}[hbt!]
    \centering
    \includegraphics[width=14.5cm]{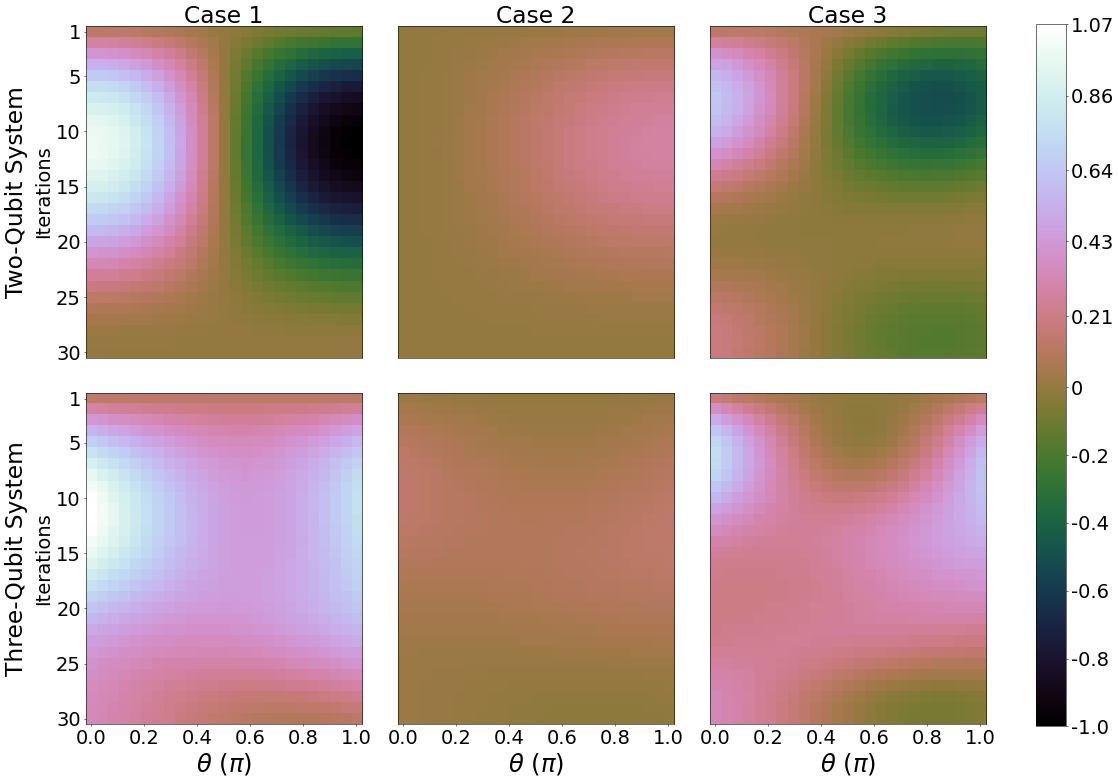}
    \caption{ {Power}
 ($P_{\Delta W_{max}}$) of the total system for $0 \ \leq \ \theta \ \leq \ \pi$ (x-axis) and $0 \ \leq \ $ iterations $ \ \leq  \ 30 $ (y-axis); First row:  Two-qubit systems where columns from left to right are cases 1, 2, and~3, respectively. Brighter shades correspond to a greater value of power. Parameters: $\epsilon_{1j} = 0\epsilon_{2B}, \ \epsilon_{2j} = 1\epsilon_{2B}$ for $j = A, B$, and~$C$. $k_{B}T_{A} = 4\epsilon_{2B}$  and $k_{B}T_{C} = 0.4\epsilon_{2B}$. $\phi = 0$. Second row: The same parameters as the first row but for three-qubit~systems.}
    \label{fig:Power}
\end{figure}
\unskip
\begin{figure}[hbt!]
    \centering
    \includegraphics[width=14.5cm]{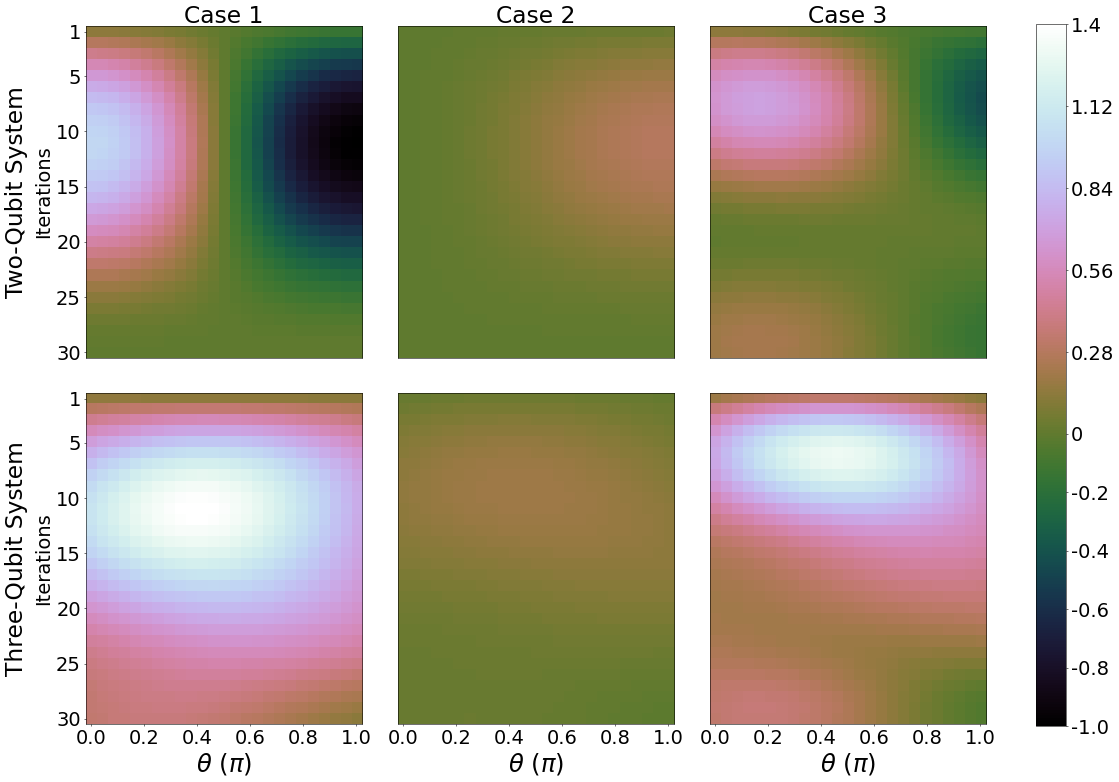}
    \caption{{The same}
 as Figure~\ref{fig:Power}, but for $\phi = \pi$.}
    \label{fig:Power2nd}
\end{figure}

Figures~\ref{fig:Power} and~\ref{fig:Power2nd} follow a somewhat similar trend to that of Figures~\ref{fig:DeltaErgTot} and~\ref{fig:DeltaErgTotPhi}, respectively (where Figures~\ref{fig:DeltaErgTot} and~\ref{fig:DeltaErgTotPhi} show $\Delta W_{max}$). Indeed,  for~the three-qubit systems, the maximum values of $P_{\Delta W_{max}}$ are reached when $\phi = \pi$ (Figure~\ref{fig:DeltaErgTotPhi}) and areas with $\theta \approx \pi/2$, where initial coherence is large. When $\phi = 0$ (Figure~\ref{fig:DeltaErgTotPhi}), the regions of high power are achieved when $\theta \approx 0 \ or \ \pi$, for~which we have lower initial coherence. These maxima appear for similar values as the maxima of  $\Delta W_{max}$, but~at earlier iterations, see \mbox{Figures~\ref{fig:DeltaErgTot} and~\ref{fig:DeltaErgTotPhi}}. Looking at Equation~(\ref{eq:powerwork}), we see that as iterations increase, a reduction in power is expected, which is what we see in Figures~\ref{fig:Power} and~\ref{fig:Power2nd}. This is why we see the power maxima at lower iterations than that of the maxima in Figures~\ref{fig:DeltaErgTot} and~\ref{fig:DeltaErgTotPhi}. Directly comparing the maxima for Figures~\ref{fig:Power} and~\ref{fig:Power2nd}, we see a larger maximum power reached when $\phi = \pi$ with a value of 1.42$\epsilon_{2B}$ compared with 1.07$\epsilon_{2B}$, giving a $32.7\%$ increase. 

\section[\appendixname~\thesubsection]{}\label{appendC}

\begin{figure}[hbt!]
    \includegraphics[width=14.5cm]{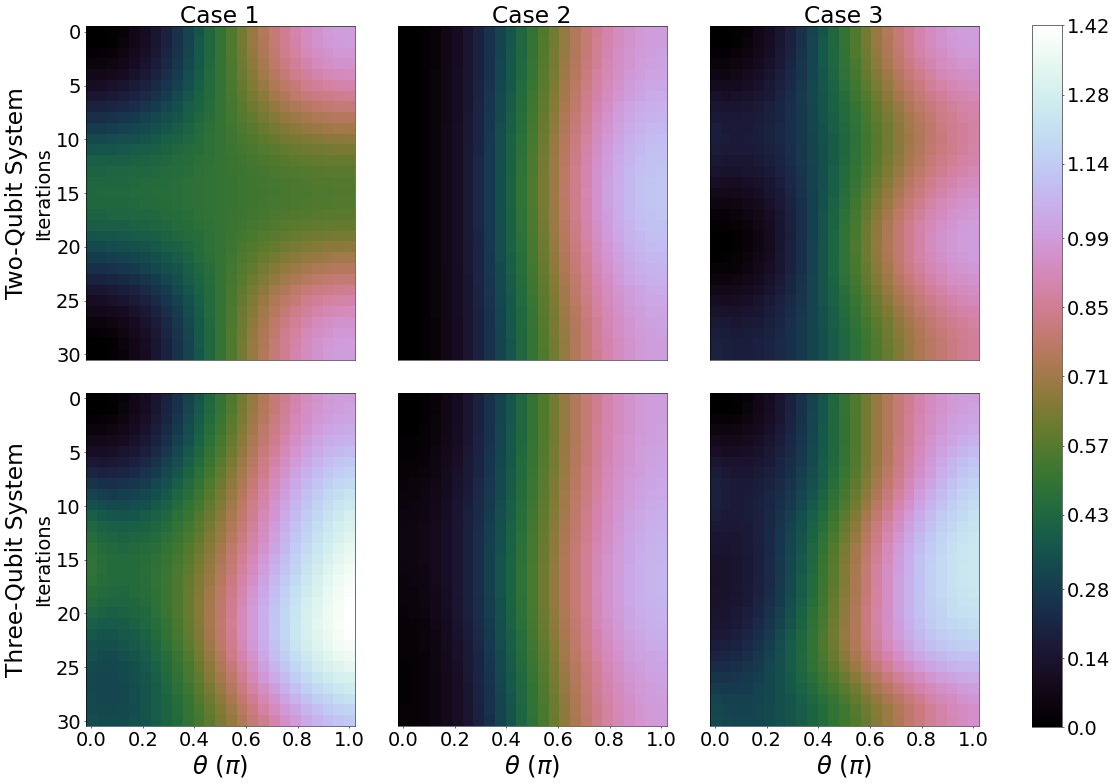}
    \caption{Ergotropy of the total system; parameters: $\epsilon_{1j} = 0\epsilon_{2B}, \ \epsilon_{2j} = 1\epsilon_{2B}$ for $j = A, B$, and~$C$. $k_{B}T_{A} = 4\epsilon_{2B}$  and $k_{B}T_{C} = 0.4\epsilon_{2B}$. $\phi = 0$. First row: Ergotropy of the total system for $0 \ \leq \ \theta \ \leq \ \pi$ (x-axis) and $0 \ \leq \ $ iterations $ \ \leq  \ 30 $ (y-axis) for two-qubit systems where columns from left to right are cases 1, 2, and~3, respectively. Brighter shades correspond to a greater value of ergotropy. Second row: The same parameters as the first row but for three-qubit~systems.}
    \label{fig:ErgoScanTotalSystem}
\end{figure}

\begin{figure}[hbt!]
    \includegraphics[width=14.5cm]{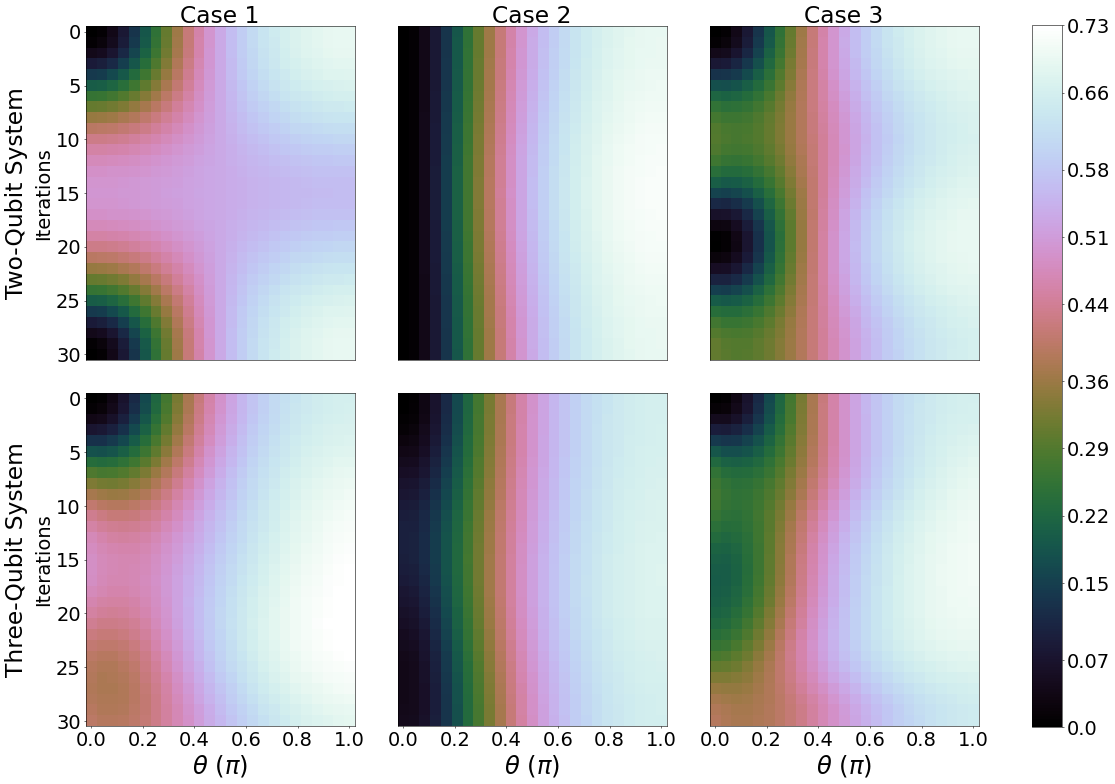}
    \caption{Ergotropy ratio $W_{ratio}$ for $\Phi = 0$, with the same parameters as Figure~\ref{fig:ErgoScanTotalSystem}. }
    \label{fig:ERGRAT}
\end{figure}

\begin{figure}[hbt!]
    \includegraphics[width=14.5cm]{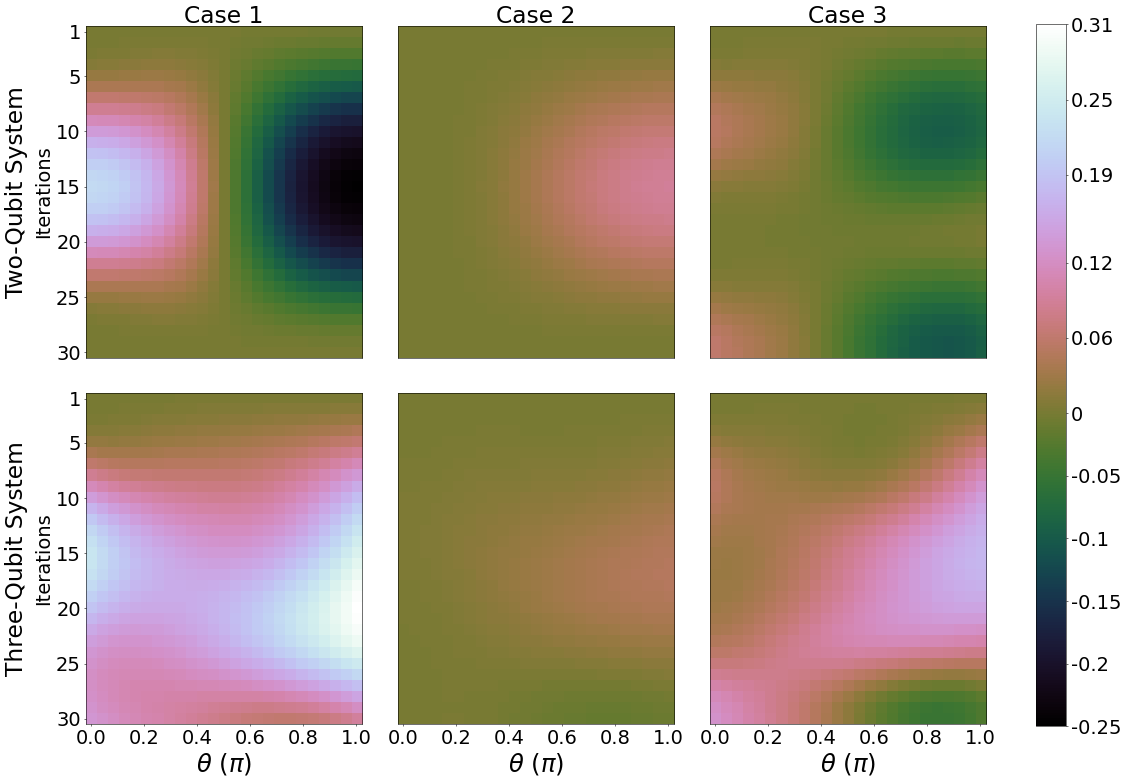}
    \caption{{Figure}
 of merit (FoM) for $\Phi = 0$, with the same parameters as Figure~\ref{fig:ErgoScanTotalSystem}.}
    \label{fig:FiguresOfMerit}
\end{figure}

For completeness, in~this Appendix, we present  results for parameter scans with qubit B initialised with  $\Phi = 0$ and for the following quantities: The ergotropy of the total system (Figure~\ref{fig:ErgoScanTotalSystem}), the ergotropy ratio $W_{ratio}$ (Figure~\ref{fig:ERGRAT}), and~the figure of merit FoM (Figure~\ref{fig:FiguresOfMerit}).

\unskip

\unskip

\clearpage

\bibliography{library.bib}

\begin{thebibliography}{48}%
\makeatletter
\providecommand \@ifxundefined [1]{%
 \@ifx{#1\undefined}
}%
\providecommand \@ifnum [1]{%
 \ifnum #1\expandafter \@firstoftwo
 \else \expandafter \@secondoftwo
 \fi
}%
\providecommand \@ifx [1]{%
 \ifx #1\expandafter \@firstoftwo
 \else \expandafter \@secondoftwo
 \fi
}%
\providecommand \natexlab [1]{#1}%
\providecommand \enquote  [1]{``#1''}%
\providecommand \bibnamefont  [1]{#1}%
\providecommand \bibfnamefont [1]{#1}%
\providecommand \citenamefont [1]{#1}%
\providecommand \href@noop [0]{\@secondoftwo}%
\providecommand \href [0]{\begingroup \@sanitize@url \@href}%
\providecommand \@href[1]{\@@startlink{#1}\@@href}%
\providecommand \@@href[1]{\endgroup#1\@@endlink}%
\providecommand \@sanitize@url [0]{\catcode `\\12\catcode `\$12\catcode `\&12\catcode `\#12\catcode `\^12\catcode `\_12\catcode `\%12\relax}%
\providecommand \@@startlink[1]{}%
\providecommand \@@endlink[0]{}%
\providecommand \url  [0]{\begingroup\@sanitize@url \@url }%
\providecommand \@url [1]{\endgroup\@href {#1}{\urlprefix }}%
\providecommand \urlprefix  [0]{URL }%
\providecommand \Eprint [0]{\href }%
\providecommand \doibase [0]{https://doi.org/}%
\providecommand \selectlanguage [0]{\@gobble}%
\providecommand \bibinfo  [0]{\@secondoftwo}%
\providecommand \bibfield  [0]{\@secondoftwo}%
\providecommand \translation [1]{[#1]}%
\providecommand \BibitemOpen [0]{}%
\providecommand \bibitemStop [0]{}%
\providecommand \bibitemNoStop [0]{.\EOS\space}%
\providecommand \EOS [0]{\spacefactor3000\relax}%
\providecommand \BibitemShut  [1]{\csname bibitem#1\endcsname}%
\let\auto@bib@innerbib\@empty
\bibitem [{\citenamefont {Ro{\ss}nagel}\ \emph {et~al.}(2016)\citenamefont {Ro{\ss}nagel}, \citenamefont {Dawkins}, \citenamefont {Tolazzi}, \citenamefont {Abah}, \citenamefont {Lutz}, \citenamefont {Schmidt-Kaler},\ and\ \citenamefont {Singer}}]{Rosnagel2016}%
  \BibitemOpen
  \bibfield  {author} {\bibinfo {author} {\bibfnamefont {J.}~\bibnamefont {Ro{\ss}nagel}}, \bibinfo {author} {\bibfnamefont {S.~T.}\ \bibnamefont {Dawkins}}, \bibinfo {author} {\bibfnamefont {K.~N.}\ \bibnamefont {Tolazzi}}, \bibinfo {author} {\bibfnamefont {O.}~\bibnamefont {Abah}}, \bibinfo {author} {\bibfnamefont {E.}~\bibnamefont {Lutz}}, \bibinfo {author} {\bibfnamefont {F.}~\bibnamefont {Schmidt-Kaler}},\ and\ \bibinfo {author} {\bibfnamefont {K.}~\bibnamefont {Singer}},\ }\bibfield  {title} {\bibinfo {title} {{A single-atom heat engine}},\ }\href {https://doi.org/10.1126/science.aad6320} {\bibfield  {journal} {\bibinfo  {journal} {Science}\ }\textbf {\bibinfo {volume} {352}},\ \bibinfo {pages} {325} (\bibinfo {year} {2016})},\ \Eprint {https://arxiv.org/abs/1510.03681} {arXiv:1510.03681} \BibitemShut {NoStop}%
\bibitem [{\citenamefont {Peterson}\ \emph {et~al.}(2019)\citenamefont {Peterson}, \citenamefont {Batalh{\~{a}}o}, \citenamefont {Herrera}, \citenamefont {Souza}, \citenamefont {Sarthour}, \citenamefont {Oliveira},\ and\ \citenamefont {Serra}}]{Peterson2019}%
  \BibitemOpen
  \bibfield  {author} {\bibinfo {author} {\bibfnamefont {J.~P.}\ \bibnamefont {Peterson}}, \bibinfo {author} {\bibfnamefont {T.~B.}\ \bibnamefont {Batalh{\~{a}}o}}, \bibinfo {author} {\bibfnamefont {M.}~\bibnamefont {Herrera}}, \bibinfo {author} {\bibfnamefont {A.~M.}\ \bibnamefont {Souza}}, \bibinfo {author} {\bibfnamefont {R.~S.}\ \bibnamefont {Sarthour}}, \bibinfo {author} {\bibfnamefont {I.~S.}\ \bibnamefont {Oliveira}},\ and\ \bibinfo {author} {\bibfnamefont {R.~M.}\ \bibnamefont {Serra}},\ }\bibfield  {title} {\bibinfo {title} {{Experimental Characterization of a Spin Quantum Heat Engine}},\ }\href {https://doi.org/10.1103/PhysRevLett.123.240601} {\bibfield  {journal} {\bibinfo  {journal} {Physical Review Letters}\ }\textbf {\bibinfo {volume} {123}},\ \bibinfo {pages} {1} (\bibinfo {year} {2019})},\ \Eprint {https://arxiv.org/abs/1803.06021} {arXiv:1803.06021} \BibitemShut {NoStop}%
\bibitem [{\citenamefont {Klatzow}\ \emph {et~al.}(2019)\citenamefont {Klatzow}, \citenamefont {Becker}, \citenamefont {Ledingham}, \citenamefont {Weinzetl}, \citenamefont {Kaczmarek}, \citenamefont {Saunders}, \citenamefont {Nunn}, \citenamefont {Walmsley}, \citenamefont {Uzdin},\ and\ \citenamefont {Poem}}]{Klatzow2019}%
  \BibitemOpen
  \bibfield  {author} {\bibinfo {author} {\bibfnamefont {J.}~\bibnamefont {Klatzow}}, \bibinfo {author} {\bibfnamefont {J.~N.}\ \bibnamefont {Becker}}, \bibinfo {author} {\bibfnamefont {P.~M.}\ \bibnamefont {Ledingham}}, \bibinfo {author} {\bibfnamefont {C.}~\bibnamefont {Weinzetl}}, \bibinfo {author} {\bibfnamefont {K.~T.}\ \bibnamefont {Kaczmarek}}, \bibinfo {author} {\bibfnamefont {D.~J.}\ \bibnamefont {Saunders}}, \bibinfo {author} {\bibfnamefont {J.}~\bibnamefont {Nunn}}, \bibinfo {author} {\bibfnamefont {I.~A.}\ \bibnamefont {Walmsley}}, \bibinfo {author} {\bibfnamefont {R.}~\bibnamefont {Uzdin}},\ and\ \bibinfo {author} {\bibfnamefont {E.}~\bibnamefont {Poem}},\ }\bibfield  {title} {\bibinfo {title} {Experimental demonstration of quantum effects in the operation of microscopic heat engines},\ }\href {https://doi.org/10.1103/PhysRevLett.122.110601} {\bibfield  {journal} {\bibinfo  {journal} {Phys. Rev. Lett.}\ }\textbf {\bibinfo {volume} {122}},\ \bibinfo {pages} {110601} (\bibinfo {year}
  {2019})}\BibitemShut {NoStop}%
\bibitem [{\citenamefont {von Lindenfels}\ \emph {et~al.}(2019)\citenamefont {von Lindenfels}, \citenamefont {Gr\"ab}, \citenamefont {Schmiegelow}, \citenamefont {Kaushal}, \citenamefont {Schulz}, \citenamefont {Mitchison}, \citenamefont {Goold}, \citenamefont {Schmidt-Kaler},\ and\ \citenamefont {Poschinger}}]{vonLindenfels2019}%
  \BibitemOpen
  \bibfield  {author} {\bibinfo {author} {\bibfnamefont {D.}~\bibnamefont {von Lindenfels}}, \bibinfo {author} {\bibfnamefont {O.}~\bibnamefont {Gr\"ab}}, \bibinfo {author} {\bibfnamefont {C.~T.}\ \bibnamefont {Schmiegelow}}, \bibinfo {author} {\bibfnamefont {V.}~\bibnamefont {Kaushal}}, \bibinfo {author} {\bibfnamefont {J.}~\bibnamefont {Schulz}}, \bibinfo {author} {\bibfnamefont {M.~T.}\ \bibnamefont {Mitchison}}, \bibinfo {author} {\bibfnamefont {J.}~\bibnamefont {Goold}}, \bibinfo {author} {\bibfnamefont {F.}~\bibnamefont {Schmidt-Kaler}},\ and\ \bibinfo {author} {\bibfnamefont {U.~G.}\ \bibnamefont {Poschinger}},\ }\bibfield  {title} {\bibinfo {title} {Spin heat engine coupled to a harmonic-oscillator flywheel},\ }\href {https://doi.org/10.1103/PhysRevLett.123.080602} {\bibfield  {journal} {\bibinfo  {journal} {Phys. Rev. Lett.}\ }\textbf {\bibinfo {volume} {123}},\ \bibinfo {pages} {080602} (\bibinfo {year} {2019})}\BibitemShut {NoStop}%
\bibitem [{\citenamefont {Van~Horne}\ \emph {et~al.}(2020)\citenamefont {Van~Horne}, \citenamefont {Yum}, \citenamefont {Dutta}, \citenamefont {Hänggi}, \citenamefont {Gong}, \citenamefont {Poletti},\ and\ \citenamefont {Mukherjee}}]{VanHorne2020}%
  \BibitemOpen
  \bibfield  {author} {\bibinfo {author} {\bibfnamefont {N.}~\bibnamefont {Van~Horne}}, \bibinfo {author} {\bibfnamefont {D.}~\bibnamefont {Yum}}, \bibinfo {author} {\bibfnamefont {T.}~\bibnamefont {Dutta}}, \bibinfo {author} {\bibfnamefont {P.}~\bibnamefont {Hänggi}}, \bibinfo {author} {\bibfnamefont {J.}~\bibnamefont {Gong}}, \bibinfo {author} {\bibfnamefont {D.}~\bibnamefont {Poletti}},\ and\ \bibinfo {author} {\bibfnamefont {M.}~\bibnamefont {Mukherjee}},\ }\bibfield  {title} {\bibinfo {title} {Single-atom energy-conversion device with a quantum load},\ }\href {https://doi.org/10.1038/s41534-020-0264-6} {\bibfield  {journal} {\bibinfo  {journal} {npj Quantum Inf}\ }\textbf {\bibinfo {volume} {6}},\ \bibinfo {pages} {37} (\bibinfo {year} {2020})}\BibitemShut {NoStop}%
\bibitem [{\citenamefont {Bouton}\ \emph {et~al.}(2021)\citenamefont {Bouton}, \citenamefont {Nettersheim}, \citenamefont {Burgardt}, \citenamefont {Adam}, \citenamefont {Lutz},\ and\ \citenamefont {Widera}}]{Bouton2021}%
  \BibitemOpen
  \bibfield  {author} {\bibinfo {author} {\bibfnamefont {Q.}~\bibnamefont {Bouton}}, \bibinfo {author} {\bibfnamefont {J.}~\bibnamefont {Nettersheim}}, \bibinfo {author} {\bibfnamefont {S.}~\bibnamefont {Burgardt}}, \bibinfo {author} {\bibfnamefont {D.}~\bibnamefont {Adam}}, \bibinfo {author} {\bibfnamefont {E.}~\bibnamefont {Lutz}},\ and\ \bibinfo {author} {\bibfnamefont {A.}~\bibnamefont {Widera}},\ }\bibfield  {title} {\bibinfo {title} {{A quantum heat engine driven by atomic collisions}},\ }\href {https://doi.org/10.1038/s41467-021-22222-z} {\bibfield  {journal} {\bibinfo  {journal} {Nature Communications}\ }\textbf {\bibinfo {volume} {12}},\ \bibinfo {pages} {1} (\bibinfo {year} {2021})}\BibitemShut {NoStop}%
\bibitem [{\citenamefont {Lisboa}\ \emph {et~al.}(2022)\citenamefont {Lisboa}, \citenamefont {Dieguez}, \citenamefont {Guimar\~aes}, \citenamefont {Santos},\ and\ \citenamefont {Serra}}]{Lisboa2022}%
  \BibitemOpen
  \bibfield  {author} {\bibinfo {author} {\bibfnamefont {V.~F.}\ \bibnamefont {Lisboa}}, \bibinfo {author} {\bibfnamefont {P.~R.}\ \bibnamefont {Dieguez}}, \bibinfo {author} {\bibfnamefont {J.~R.}\ \bibnamefont {Guimar\~aes}}, \bibinfo {author} {\bibfnamefont {J.~F.~G.}\ \bibnamefont {Santos}},\ and\ \bibinfo {author} {\bibfnamefont {R.~M.}\ \bibnamefont {Serra}},\ }\bibfield  {title} {\bibinfo {title} {Experimental investigation of a quantum heat engine powered by generalized measurements},\ }\href {https://doi.org/10.1103/PhysRevA.106.022436} {\bibfield  {journal} {\bibinfo  {journal} {Phys. Rev. A}\ }\textbf {\bibinfo {volume} {106}},\ \bibinfo {pages} {022436} (\bibinfo {year} {2022})}\BibitemShut {NoStop}%
\bibitem [{\citenamefont {Herrera}\ \emph {et~al.}(2023)\citenamefont {Herrera}, \citenamefont {Reina}, \citenamefont {D'Amico},\ and\ \citenamefont {Serra}}]{Herrera2023}%
  \BibitemOpen
  \bibfield  {author} {\bibinfo {author} {\bibfnamefont {M.}~\bibnamefont {Herrera}}, \bibinfo {author} {\bibfnamefont {J.~H.}\ \bibnamefont {Reina}}, \bibinfo {author} {\bibfnamefont {I.}~\bibnamefont {D'Amico}},\ and\ \bibinfo {author} {\bibfnamefont {R.~M.}\ \bibnamefont {Serra}},\ }\bibfield  {title} {\bibinfo {title} {Correlation-boosted quantum engine: A proof-of-principle demonstration},\ }\href {https://doi.org/10.1103/PhysRevResearch.5.043104} {\bibfield  {journal} {\bibinfo  {journal} {Phys. Rev. Res.}\ }\textbf {\bibinfo {volume} {5}},\ \bibinfo {pages} {043104} (\bibinfo {year} {2023})}\BibitemShut {NoStop}%
\bibitem [{\citenamefont {Dillenschneider}\ and\ \citenamefont {Lutz}(2009)}]{Dillenschneider2009}%
  \BibitemOpen
  \bibfield  {author} {\bibinfo {author} {\bibfnamefont {R.}~\bibnamefont {Dillenschneider}}\ and\ \bibinfo {author} {\bibfnamefont {E.}~\bibnamefont {Lutz}},\ }\bibfield  {title} {\bibinfo {title} {{Energetics of quantum correlations}},\ }\href {https://doi.org/10.1209/0295-5075/88/50003} {\bibfield  {journal} {\bibinfo  {journal} {Epl}\ }\textbf {\bibinfo {volume} {88}},\ \bibinfo {pages} {50003} (\bibinfo {year} {2009})},\ \Eprint {https://arxiv.org/abs/0803.4067} {arXiv:0803.4067} \BibitemShut {NoStop}%
\bibitem [{\citenamefont {Ro\ss{}nagel}\ \emph {et~al.}(2014)\citenamefont {Ro\ss{}nagel}, \citenamefont {Abah}, \citenamefont {Schmidt-Kaler}, \citenamefont {Singer},\ and\ \citenamefont {Lutz}}]{Rossnagel2014}%
  \BibitemOpen
  \bibfield  {author} {\bibinfo {author} {\bibfnamefont {J.}~\bibnamefont {Ro\ss{}nagel}}, \bibinfo {author} {\bibfnamefont {O.}~\bibnamefont {Abah}}, \bibinfo {author} {\bibfnamefont {F.}~\bibnamefont {Schmidt-Kaler}}, \bibinfo {author} {\bibfnamefont {K.}~\bibnamefont {Singer}},\ and\ \bibinfo {author} {\bibfnamefont {E.}~\bibnamefont {Lutz}},\ }\bibfield  {title} {\bibinfo {title} {Nanoscale heat engine beyond the carnot limit},\ }\href {https://doi.org/10.1103/PhysRevLett.112.030602} {\bibfield  {journal} {\bibinfo  {journal} {Phys. Rev. Lett.}\ }\textbf {\bibinfo {volume} {112}},\ \bibinfo {pages} {030602} (\bibinfo {year} {2014})}\BibitemShut {NoStop}%
\bibitem [{\citenamefont {Vinjanampathy}\ and\ \citenamefont {Anders}(2016)}]{Vinjanampathy2016}%
  \BibitemOpen
  \bibfield  {author} {\bibinfo {author} {\bibfnamefont {S.}~\bibnamefont {Vinjanampathy}}\ and\ \bibinfo {author} {\bibfnamefont {J.}~\bibnamefont {Anders}},\ }\bibfield  {title} {\bibinfo {title} {Quantum thermodynamics},\ }\href {https://doi.org/10.1080/00107514.2016.1201896} {\bibfield  {journal} {\bibinfo  {journal} {Contemporary Physics}\ }\textbf {\bibinfo {volume} {57}},\ \bibinfo {pages} {545} (\bibinfo {year} {2016})}\BibitemShut {NoStop}%
\bibitem [{\citenamefont {Bera}\ \emph {et~al.}(2017)\citenamefont {Bera}, \citenamefont {Riera}, \citenamefont {Lewenstein},\ and\ \citenamefont {Winter}}]{Bera2017}%
  \BibitemOpen
  \bibfield  {author} {\bibinfo {author} {\bibfnamefont {M.~N.}\ \bibnamefont {Bera}}, \bibinfo {author} {\bibfnamefont {A.}~\bibnamefont {Riera}}, \bibinfo {author} {\bibfnamefont {M.}~\bibnamefont {Lewenstein}},\ and\ \bibinfo {author} {\bibfnamefont {A.}~\bibnamefont {Winter}},\ }\bibfield  {title} {\bibinfo {title} {Generalized laws of thermodynamics in the presence of correlations},\ }\href {https://doi.org/10.1038/s41467-017-02370-x} {\bibfield  {journal} {\bibinfo  {journal} {Nature Communications}\ }\textbf {\bibinfo {volume} {8}},\ \bibinfo {pages} {2180} (\bibinfo {year} {2017})}\BibitemShut {NoStop}%
\bibitem [{\citenamefont {Campaioli}\ \emph {et~al.}(2017)\citenamefont {Campaioli}, \citenamefont {Pollock}, \citenamefont {Binder}, \citenamefont {Céleri}, \citenamefont {Goold}, \citenamefont {Vinjanampathy},\ and\ \citenamefont {Modi}}]{Campaioli2017}%
  \BibitemOpen
  \bibfield  {author} {\bibinfo {author} {\bibfnamefont {F.}~\bibnamefont {Campaioli}}, \bibinfo {author} {\bibfnamefont {F.~A.}\ \bibnamefont {Pollock}}, \bibinfo {author} {\bibfnamefont {F.~C.}\ \bibnamefont {Binder}}, \bibinfo {author} {\bibfnamefont {L.}~\bibnamefont {Céleri}}, \bibinfo {author} {\bibfnamefont {J.}~\bibnamefont {Goold}}, \bibinfo {author} {\bibfnamefont {S.}~\bibnamefont {Vinjanampathy}},\ and\ \bibinfo {author} {\bibfnamefont {K.}~\bibnamefont {Modi}},\ }\bibfield  {title} {\bibinfo {title} {Enhancing the charging power of quantum batteries},\ }\href {https://doi.org/10.1103/PhysRevLett.118.150601} {\bibfield  {journal} {\bibinfo  {journal} {Physical Review Letters}\ }\textbf {\bibinfo {volume} {118}},\ \bibinfo {pages} {150601} (\bibinfo {year} {2017})}\BibitemShut {NoStop}%
\bibitem [{\citenamefont {Ferraro}\ \emph {et~al.}(2018)\citenamefont {Ferraro}, \citenamefont {Campisi}, \citenamefont {Andolina}, \citenamefont {Pellegrini},\ and\ \citenamefont {Polini}}]{Ferraro2018}%
  \BibitemOpen
  \bibfield  {author} {\bibinfo {author} {\bibfnamefont {D.}~\bibnamefont {Ferraro}}, \bibinfo {author} {\bibfnamefont {M.}~\bibnamefont {Campisi}}, \bibinfo {author} {\bibfnamefont {G.~M.}\ \bibnamefont {Andolina}}, \bibinfo {author} {\bibfnamefont {V.}~\bibnamefont {Pellegrini}},\ and\ \bibinfo {author} {\bibfnamefont {M.}~\bibnamefont {Polini}},\ }\bibfield  {title} {\bibinfo {title} {High-power collective charging of a solid-state quantum battery},\ }\href {https://doi.org/10.1103/PhysRevLett.120.117702} {\bibfield  {journal} {\bibinfo  {journal} {Phys. Rev. Lett.}\ }\textbf {\bibinfo {volume} {120}},\ \bibinfo {pages} {117702} (\bibinfo {year} {2018})}\BibitemShut {NoStop}%
\bibitem [{\citenamefont {Le}\ \emph {et~al.}(2018)\citenamefont {Le}, \citenamefont {Levinsen}, \citenamefont {Modi}, \citenamefont {Parish},\ and\ \citenamefont {Pollock}}]{Le2018}%
  \BibitemOpen
  \bibfield  {author} {\bibinfo {author} {\bibfnamefont {T.~P.}\ \bibnamefont {Le}}, \bibinfo {author} {\bibfnamefont {J.}~\bibnamefont {Levinsen}}, \bibinfo {author} {\bibfnamefont {K.}~\bibnamefont {Modi}}, \bibinfo {author} {\bibfnamefont {M.~M.}\ \bibnamefont {Parish}},\ and\ \bibinfo {author} {\bibfnamefont {F.~A.}\ \bibnamefont {Pollock}},\ }\bibfield  {title} {\bibinfo {title} {Spin-chain model of a many-body quantum battery},\ }\href {https://doi.org/10.1103/PhysRevA.97.022106} {\bibfield  {journal} {\bibinfo  {journal} {Phys. Rev. A}\ }\textbf {\bibinfo {volume} {97}},\ \bibinfo {pages} {022106} (\bibinfo {year} {2018})}\BibitemShut {NoStop}%
\bibitem [{\citenamefont {Zhang}\ \emph {et~al.}(2019)\citenamefont {Zhang}, \citenamefont {Yang}, \citenamefont {Fu},\ and\ \citenamefont {Wang}}]{Zhang2018}%
  \BibitemOpen
  \bibfield  {author} {\bibinfo {author} {\bibfnamefont {Y.-Y.}\ \bibnamefont {Zhang}}, \bibinfo {author} {\bibfnamefont {T.-R.}\ \bibnamefont {Yang}}, \bibinfo {author} {\bibfnamefont {L.}~\bibnamefont {Fu}},\ and\ \bibinfo {author} {\bibfnamefont {X.}~\bibnamefont {Wang}},\ }\bibfield  {title} {\bibinfo {title} {Powerful harmonic charging in a quantum battery},\ }\href {https://doi.org/10.1103/PhysRevE.99.052106} {\bibfield  {journal} {\bibinfo  {journal} {Phys. Rev. E}\ }\textbf {\bibinfo {volume} {99}},\ \bibinfo {pages} {052106} (\bibinfo {year} {2019})}\BibitemShut {NoStop}%
\bibitem [{\citenamefont {Joshi}\ and\ \citenamefont {Mahesh}(2022)}]{Joshi2022}%
  \BibitemOpen
  \bibfield  {author} {\bibinfo {author} {\bibfnamefont {J.}~\bibnamefont {Joshi}}\ and\ \bibinfo {author} {\bibfnamefont {T.~S.}\ \bibnamefont {Mahesh}},\ }\bibfield  {title} {\bibinfo {title} {Experimental investigation of a quantum battery using star-topology nmr spin systems},\ }\href {https://doi.org/10.1103/PhysRevA.106.042601} {\bibfield  {journal} {\bibinfo  {journal} {Phys. Rev. A}\ }\textbf {\bibinfo {volume} {106}},\ \bibinfo {pages} {042601} (\bibinfo {year} {2022})}\BibitemShut {NoStop}%
\bibitem [{\citenamefont {Yao}\ and\ \citenamefont {Shao}(2022)}]{Yao2022}%
  \BibitemOpen
  \bibfield  {author} {\bibinfo {author} {\bibfnamefont {Y.}~\bibnamefont {Yao}}\ and\ \bibinfo {author} {\bibfnamefont {X.~Q.}\ \bibnamefont {Shao}},\ }\bibfield  {title} {\bibinfo {title} {Optimal charging of open spin-chain quantum batteries via homodyne-based feedback control},\ }\href {https://doi.org/10.1103/PhysRevE.106.014138} {\bibfield  {journal} {\bibinfo  {journal} {Physical Review E}\ }\textbf {\bibinfo {volume} {106}},\ \bibinfo {pages} {014138} (\bibinfo {year} {2022})}\BibitemShut {NoStop}%
\bibitem [{\citenamefont {Binder}\ \emph {et~al.}(2015)\citenamefont {Binder}, \citenamefont {Vinjanampathy}, \citenamefont {Modi},\ and\ \citenamefont {Goold}}]{Binder2015}%
  \BibitemOpen
  \bibfield  {author} {\bibinfo {author} {\bibfnamefont {F.~C.}\ \bibnamefont {Binder}}, \bibinfo {author} {\bibfnamefont {S.}~\bibnamefont {Vinjanampathy}}, \bibinfo {author} {\bibfnamefont {K.}~\bibnamefont {Modi}},\ and\ \bibinfo {author} {\bibfnamefont {J.}~\bibnamefont {Goold}},\ }\bibfield  {title} {\bibinfo {title} {Quantacell: Powerful charging of quantum batteries},\ }\href {https://doi.org/10.1088/1367-2630/17/7/075015} {\bibfield  {journal} {\bibinfo  {journal} {New Journal of Physics}\ }\textbf {\bibinfo {volume} {17}},\ \bibinfo {pages} {075015} (\bibinfo {year} {2015})}\BibitemShut {NoStop}%
\bibitem [{\citenamefont {Campaioli}\ \emph {et~al.}(2018)\citenamefont {Campaioli}, \citenamefont {Pollock},\ and\ \citenamefont {Vinjanampathy}}]{Campaioli2018}%
  \BibitemOpen
  \bibfield  {author} {\bibinfo {author} {\bibfnamefont {F.}~\bibnamefont {Campaioli}}, \bibinfo {author} {\bibfnamefont {F.~A.}\ \bibnamefont {Pollock}},\ and\ \bibinfo {author} {\bibfnamefont {S.}~\bibnamefont {Vinjanampathy}},\ }\href {https://doi.org/10.48550/arXiv.1805.05507} {\emph {\bibinfo {title} {Quantum Batteries - Review Chapter}}}\ (\bibinfo  {publisher} {Springer},\ \bibinfo {year} {2018})\BibitemShut {NoStop}%
\bibitem [{\citenamefont {Alicki}\ and\ \citenamefont {Fannes}(2013)}]{Alicki2013}%
  \BibitemOpen
  \bibfield  {author} {\bibinfo {author} {\bibfnamefont {R.}~\bibnamefont {Alicki}}\ and\ \bibinfo {author} {\bibfnamefont {M.}~\bibnamefont {Fannes}},\ }\bibfield  {title} {\bibinfo {title} {Entanglement boost for extractable work from ensembles of quantum batteries},\ }\href {https://doi.org/10.1103/PhysRevE.87.042123} {\bibfield  {journal} {\bibinfo  {journal} {Physical Review E - Statistical, Nonlinear, and Soft Matter Physics}\ }\textbf {\bibinfo {volume} {87}},\ \bibinfo {pages} {042123} (\bibinfo {year} {2013})}\BibitemShut {NoStop}%
\bibitem [{\citenamefont {Pijn}\ \emph {et~al.}(2022)\citenamefont {Pijn}, \citenamefont {Onishchenko}, \citenamefont {Hilder}, \citenamefont {Poschinger}, \citenamefont {Schmidt-Kaler},\ and\ \citenamefont {Uzdin}}]{PhysRevLett.128.110601}%
  \BibitemOpen
  \bibfield  {author} {\bibinfo {author} {\bibfnamefont {D.}~\bibnamefont {Pijn}}, \bibinfo {author} {\bibfnamefont {O.}~\bibnamefont {Onishchenko}}, \bibinfo {author} {\bibfnamefont {J.}~\bibnamefont {Hilder}}, \bibinfo {author} {\bibfnamefont {U.~G.}\ \bibnamefont {Poschinger}}, \bibinfo {author} {\bibfnamefont {F.}~\bibnamefont {Schmidt-Kaler}},\ and\ \bibinfo {author} {\bibfnamefont {R.}~\bibnamefont {Uzdin}},\ }\bibfield  {title} {\bibinfo {title} {Detecting heat leaks with trapped ion qubits},\ }\href {https://doi.org/10.1103/PhysRevLett.128.110601} {\bibfield  {journal} {\bibinfo  {journal} {Phys. Rev. Lett.}\ }\textbf {\bibinfo {volume} {128}},\ \bibinfo {pages} {110601} (\bibinfo {year} {2022})}\BibitemShut {NoStop}%
\bibitem [{\citenamefont {Wu}\ \emph {et~al.}(2021)\citenamefont {Wu}, \citenamefont {Liang}, \citenamefont {Tian}, \citenamefont {Yang}, \citenamefont {Chen}, \citenamefont {Liu}, \citenamefont {Tey},\ and\ \citenamefont {You}}]{Wu2021}%
  \BibitemOpen
  \bibfield  {author} {\bibinfo {author} {\bibfnamefont {X.}~\bibnamefont {Wu}}, \bibinfo {author} {\bibfnamefont {X.}~\bibnamefont {Liang}}, \bibinfo {author} {\bibfnamefont {Y.}~\bibnamefont {Tian}}, \bibinfo {author} {\bibfnamefont {F.}~\bibnamefont {Yang}}, \bibinfo {author} {\bibfnamefont {C.}~\bibnamefont {Chen}}, \bibinfo {author} {\bibfnamefont {Y.~C.}\ \bibnamefont {Liu}}, \bibinfo {author} {\bibfnamefont {M.~K.}\ \bibnamefont {Tey}},\ and\ \bibinfo {author} {\bibfnamefont {L.}~\bibnamefont {You}},\ }\bibfield  {title} {\bibinfo {title} {A concise review of rydberg atom based quantum computation and quantum simulation},\ }\href {https://doi.org/10.1088/1674-1056/abd76f} {\bibfield  {journal} {\bibinfo  {journal} {Chinese Physics B}\ }\textbf {\bibinfo {volume} {30}},\ \bibinfo {pages} {020305} (\bibinfo {year} {2021})}\BibitemShut {NoStop}%
\bibitem [{\citenamefont {Abdelhafez}\ \emph {et~al.}(2020)\citenamefont {Abdelhafez}, \citenamefont {Baker}, \citenamefont {Gyenis}, \citenamefont {Mundada}, \citenamefont {Houck}, \citenamefont {Schuster},\ and\ \citenamefont {Koch}}]{Abdelhafez2020}%
  \BibitemOpen
  \bibfield  {author} {\bibinfo {author} {\bibfnamefont {M.}~\bibnamefont {Abdelhafez}}, \bibinfo {author} {\bibfnamefont {B.}~\bibnamefont {Baker}}, \bibinfo {author} {\bibfnamefont {A.}~\bibnamefont {Gyenis}}, \bibinfo {author} {\bibfnamefont {P.}~\bibnamefont {Mundada}}, \bibinfo {author} {\bibfnamefont {A.~A.}\ \bibnamefont {Houck}}, \bibinfo {author} {\bibfnamefont {D.}~\bibnamefont {Schuster}},\ and\ \bibinfo {author} {\bibfnamefont {J.}~\bibnamefont {Koch}},\ }\bibfield  {title} {\bibinfo {title} {Universal gates for protected superconducting qubits using optimal control},\ }\href {https://doi.org/10.1103/PhysRevA.101.022321} {\bibfield  {journal} {\bibinfo  {journal} {Phys. Rev. A}\ }\textbf {\bibinfo {volume} {101}},\ \bibinfo {pages} {022321} (\bibinfo {year} {2020})}\BibitemShut {NoStop}%
\bibitem [{\citenamefont {Giorgi}\ and\ \citenamefont {Campbell}(2015)}]{Giorgi2015}%
  \BibitemOpen
  \bibfield  {author} {\bibinfo {author} {\bibfnamefont {G.~L.}\ \bibnamefont {Giorgi}}\ and\ \bibinfo {author} {\bibfnamefont {S.}~\bibnamefont {Campbell}},\ }\bibfield  {title} {\bibinfo {title} {Correlation approach to work extraction from finite quantum systems},\ }\href {https://doi.org/10.1088/0953-4075/48/3/035501} {\bibfield  {journal} {\bibinfo  {journal} {Journal of Physics B: Atomic, Molecular and Optical Physics}\ }\textbf {\bibinfo {volume} {48}},\ \bibinfo {pages} {035501} (\bibinfo {year} {2015})}\BibitemShut {NoStop}%
\bibitem [{\citenamefont {Perarnau-Llobet}\ \emph {et~al.}(2015)\citenamefont {Perarnau-Llobet}, \citenamefont {Hovhannisyan}, \citenamefont {Huber}, \citenamefont {Skrzypczyk}, \citenamefont {Tura},\ and\ \citenamefont {Ac\'{\i}n}}]{Perarnau2015}%
  \BibitemOpen
  \bibfield  {author} {\bibinfo {author} {\bibfnamefont {M.}~\bibnamefont {Perarnau-Llobet}}, \bibinfo {author} {\bibfnamefont {K.~V.}\ \bibnamefont {Hovhannisyan}}, \bibinfo {author} {\bibfnamefont {M.}~\bibnamefont {Huber}}, \bibinfo {author} {\bibfnamefont {P.}~\bibnamefont {Skrzypczyk}}, \bibinfo {author} {\bibfnamefont {J.}~\bibnamefont {Tura}},\ and\ \bibinfo {author} {\bibfnamefont {A.}~\bibnamefont {Ac\'{\i}n}},\ }\bibfield  {title} {\bibinfo {title} {Most energetic passive states},\ }\href {https://doi.org/10.1103/PhysRevE.92.042147} {\bibfield  {journal} {\bibinfo  {journal} {Phys. Rev. E}\ }\textbf {\bibinfo {volume} {92}},\ \bibinfo {pages} {042147} (\bibinfo {year} {2015})}\BibitemShut {NoStop}%
\bibitem [{\citenamefont {Shaghaghi}\ \emph {et~al.}(2022)\citenamefont {Shaghaghi}, \citenamefont {Singh}, \citenamefont {Benenti},\ and\ \citenamefont {Rosa}}]{Shaghaghi_2022}%
  \BibitemOpen
  \bibfield  {author} {\bibinfo {author} {\bibfnamefont {V.}~\bibnamefont {Shaghaghi}}, \bibinfo {author} {\bibfnamefont {V.}~\bibnamefont {Singh}}, \bibinfo {author} {\bibfnamefont {G.}~\bibnamefont {Benenti}},\ and\ \bibinfo {author} {\bibfnamefont {D.}~\bibnamefont {Rosa}},\ }\bibfield  {title} {\bibinfo {title} {Micromasers as quantum batteries},\ }\href {https://doi.org/10.1088/2058-9565/ac8829} {\bibfield  {journal} {\bibinfo  {journal} {Quantum Science and Technology}\ }\textbf {\bibinfo {volume} {7}},\ \bibinfo {pages} {04LT01} (\bibinfo {year} {2022})}\BibitemShut {NoStop}%
\bibitem [{\citenamefont {Miller}\ \emph {et~al.}(2019)\citenamefont {Miller}, \citenamefont {Scandi}, \citenamefont {Anders},\ and\ \citenamefont {Perarnau-Llobet}}]{Miller2019}%
  \BibitemOpen
  \bibfield  {author} {\bibinfo {author} {\bibfnamefont {H.~J.~D.}\ \bibnamefont {Miller}}, \bibinfo {author} {\bibfnamefont {M.}~\bibnamefont {Scandi}}, \bibinfo {author} {\bibfnamefont {J.}~\bibnamefont {Anders}},\ and\ \bibinfo {author} {\bibfnamefont {M.}~\bibnamefont {Perarnau-Llobet}},\ }\bibfield  {title} {\bibinfo {title} {Work fluctuations in slow processes: Quantum signatures and optimal control},\ }\href {https://doi.org/10.1103/PhysRevLett.123.230603} {\bibfield  {journal} {\bibinfo  {journal} {Phys. Rev. Lett.}\ }\textbf {\bibinfo {volume} {123}},\ \bibinfo {pages} {230603} (\bibinfo {year} {2019})}\BibitemShut {NoStop}%
\bibitem [{\citenamefont {Scandi}\ \emph {et~al.}(2020)\citenamefont {Scandi}, \citenamefont {Miller}, \citenamefont {Anders},\ and\ \citenamefont {Perarnau-Llobet}}]{Scandi2020}%
  \BibitemOpen
  \bibfield  {author} {\bibinfo {author} {\bibfnamefont {M.}~\bibnamefont {Scandi}}, \bibinfo {author} {\bibfnamefont {H.~J.~D.}\ \bibnamefont {Miller}}, \bibinfo {author} {\bibfnamefont {J.}~\bibnamefont {Anders}},\ and\ \bibinfo {author} {\bibfnamefont {M.}~\bibnamefont {Perarnau-Llobet}},\ }\bibfield  {title} {\bibinfo {title} {Quantum work statistics close to equilibrium},\ }\href {https://doi.org/10.1103/PhysRevResearch.2.023377} {\bibfield  {journal} {\bibinfo  {journal} {Phys. Rev. Res.}\ }\textbf {\bibinfo {volume} {2}},\ \bibinfo {pages} {023377} (\bibinfo {year} {2020})}\BibitemShut {NoStop}%
\bibitem [{\citenamefont {Onishchenko}\ \emph {et~al.}(2024)\citenamefont {Onishchenko}, \citenamefont {Guarnieri}, \citenamefont {Rosillo-Rodes}, \citenamefont {Pijn}, \citenamefont {Hilder}, \citenamefont {Poschinger}, \citenamefont {Perarnau-Llobet}, \citenamefont {Eisert},\ and\ \citenamefont {Schmidt-Kaler}}]{Onishenko2024}%
  \BibitemOpen
  \bibfield  {author} {\bibinfo {author} {\bibfnamefont {O.}~\bibnamefont {Onishchenko}}, \bibinfo {author} {\bibfnamefont {G.}~\bibnamefont {Guarnieri}}, \bibinfo {author} {\bibfnamefont {P.}~\bibnamefont {Rosillo-Rodes}}, \bibinfo {author} {\bibfnamefont {D.}~\bibnamefont {Pijn}}, \bibinfo {author} {\bibfnamefont {J.}~\bibnamefont {Hilder}}, \bibinfo {author} {\bibfnamefont {U.~G.}\ \bibnamefont {Poschinger}}, \bibinfo {author} {\bibfnamefont {M.}~\bibnamefont {Perarnau-Llobet}}, \bibinfo {author} {\bibfnamefont {J.}~\bibnamefont {Eisert}},\ and\ \bibinfo {author} {\bibfnamefont {F.}~\bibnamefont {Schmidt-Kaler}},\ }\bibfield  {title} {\bibinfo {title} {Probing coherent quantum thermodynamics using a trapped ion},\ }\href {https://www.nature.com/articles/s41467-024-51263-3} {\bibfield  {journal} {\bibinfo  {journal} {Nat. Comm.}\ }\textbf {\bibinfo {volume} {15}},\ \bibinfo {pages} {6974} (\bibinfo {year} {2024})}\BibitemShut {NoStop}%
\bibitem [{\citenamefont {Morrone}\ \emph {et~al.}(2023)\citenamefont {Morrone}, \citenamefont {Rossi}, \citenamefont {Smirne},\ and\ \citenamefont {Genoni}}]{Morrone_2023}%
  \BibitemOpen
  \bibfield  {author} {\bibinfo {author} {\bibfnamefont {D.}~\bibnamefont {Morrone}}, \bibinfo {author} {\bibfnamefont {M.~A.~C.}\ \bibnamefont {Rossi}}, \bibinfo {author} {\bibfnamefont {A.}~\bibnamefont {Smirne}},\ and\ \bibinfo {author} {\bibfnamefont {M.~G.}\ \bibnamefont {Genoni}},\ }\bibfield  {title} {\bibinfo {title} {Charging a quantum battery in a non-markovian environment: a collisional model approach},\ }\href {https://doi.org/10.1088/2058-9565/accca4} {\bibfield  {journal} {\bibinfo  {journal} {Quantum Science and Technology}\ }\textbf {\bibinfo {volume} {8}},\ \bibinfo {pages} {035007} (\bibinfo {year} {2023})}\BibitemShut {NoStop}%
\bibitem [{\citenamefont {Landi}(2021)}]{e23121627}%
  \BibitemOpen
  \bibfield  {author} {\bibinfo {author} {\bibfnamefont {G.~T.}\ \bibnamefont {Landi}},\ }\bibfield  {title} {\bibinfo {title} {Battery charging in collision models with bayesian risk strategies},\ }\bibfield  {journal} {\bibinfo  {journal} {Entropy}\ }\textbf {\bibinfo {volume} {23}},\ \href {https://doi.org/10.3390/e23121627} {10.3390/e23121627} (\bibinfo {year} {2021})\BibitemShut {NoStop}%
\bibitem [{\citenamefont {Barra}(2022)}]{e24060820}%
  \BibitemOpen
  \bibfield  {author} {\bibinfo {author} {\bibfnamefont {F.}~\bibnamefont {Barra}},\ }\bibfield  {title} {\bibinfo {title} {Efficiency fluctuations in a quantum battery charged by a repeated interaction process},\ }\bibfield  {journal} {\bibinfo  {journal} {Entropy}\ }\textbf {\bibinfo {volume} {24}},\ \href {https://doi.org/10.3390/e24060820} {10.3390/e24060820} (\bibinfo {year} {2022})\BibitemShut {NoStop}%
\bibitem [{\citenamefont {Lenard}(1978)}]{Lenard1978}%
  \BibitemOpen
  \bibfield  {author} {\bibinfo {author} {\bibfnamefont {A.}~\bibnamefont {Lenard}},\ }\bibfield  {title} {\bibinfo {title} {Thermodynamical proof of the gibbs formula for elementary quantum systems},\ }\href@noop {} {\bibfield  {journal} {\bibinfo  {journal} {Journal of Statistical Physics}\ }\textbf {\bibinfo {volume} {19}} (\bibinfo {year} {1978})}\BibitemShut {NoStop}%
\bibitem [{\citenamefont {Gyhm}\ and\ \citenamefont {Fischer}(2024)}]{10.1116/5.0184903}%
  \BibitemOpen
  \bibfield  {author} {\bibinfo {author} {\bibfnamefont {J.-Y.}\ \bibnamefont {Gyhm}}\ and\ \bibinfo {author} {\bibfnamefont {U.~R.}\ \bibnamefont {Fischer}},\ }\bibfield  {title} {\bibinfo {title} {{Beneficial and detrimental entanglement for quantum battery charging}},\ }\href {https://doi.org/10.1116/5.0184903} {\bibfield  {journal} {\bibinfo  {journal} {AVS Quantum Science}\ }\textbf {\bibinfo {volume} {6}},\ \bibinfo {pages} {012001} (\bibinfo {year} {2024})},\ \Eprint {https://arxiv.org/abs/https://pubs.aip.org/avs/aqs/article-pdf/doi/10.1116/5.0184903/18703178/012001\_1\_5.0184903.pdf} {https://pubs.aip.org/avs/aqs/article-pdf/doi/10.1116/5.0184903/18703178/012001\_1\_5.0184903.pdf} \BibitemShut {NoStop}%
\bibitem [{\citenamefont {{Nikolov}}(2019)}]{Nikolov}%
  \BibitemOpen
  \bibfield  {author} {\bibinfo {author} {\bibfnamefont {P.}~\bibnamefont {{Nikolov}}},\ }\bibfield  {title} {\bibinfo {title} {{Controlled nth root of X gate on a real quantum computer}},\ }in\ \href {https://doi.org/10.1063/1.5133583} {\emph {\bibinfo {booktitle} {45th International Conference on Application of Mathematics in Engineering and Economics (AMEE'19)}}},\ \bibinfo {series} {American Institute of Physics Conference Series}, Vol.\ \bibinfo {volume} {2172}\ (\bibinfo {year} {2019})\ p.\ \bibinfo {pages} {090006}\BibitemShut {NoStop}%
\bibitem [{\citenamefont {Allahverdyan}\ \emph {et~al.}(2004)\citenamefont {Allahverdyan}, \citenamefont {Balian},\ and\ \citenamefont {Nieuwenhuizen}}]{Allahverdyan2004}%
  \BibitemOpen
  \bibfield  {author} {\bibinfo {author} {\bibfnamefont {A.~E.}\ \bibnamefont {Allahverdyan}}, \bibinfo {author} {\bibfnamefont {R.}~\bibnamefont {Balian}},\ and\ \bibinfo {author} {\bibfnamefont {T.~M.}\ \bibnamefont {Nieuwenhuizen}},\ }\bibfield  {title} {\bibinfo {title} {Maximal work extraction from finite quantum systems},\ }\href {https://doi.org/10.1209/epl/i2004-10101-2} {\bibfield  {journal} {\bibinfo  {journal} {Europhysics Letters}\ }\textbf {\bibinfo {volume} {67}},\ \bibinfo {pages} {565} (\bibinfo {year} {2004})}\BibitemShut {NoStop}%
\bibitem [{\citenamefont {Pusz}\ and\ \citenamefont {Woronowicz}(1978)}]{Pusz1978}%
  \BibitemOpen
  \bibfield  {author} {\bibinfo {author} {\bibfnamefont {W.}~\bibnamefont {Pusz}}\ and\ \bibinfo {author} {\bibfnamefont {S.~L.}\ \bibnamefont {Woronowicz}},\ }\bibfield  {title} {\bibinfo {title} {Passive states and kms states for general quantum systems},\ }\href@noop {} {\bibfield  {journal} {\bibinfo  {journal} {Commun. math. Phys}\ }\textbf {\bibinfo {volume} {58}},\ \bibinfo {pages} {273} (\bibinfo {year} {1978})}\BibitemShut {NoStop}%
\bibitem [{\citenamefont {Andolina}\ \emph {et~al.}(2019)\citenamefont {Andolina}, \citenamefont {Keck}, \citenamefont {Mari}, \citenamefont {Campisi}, \citenamefont {Giovannetti},\ and\ \citenamefont {Polini}}]{PhysRevLett.122.047702}%
  \BibitemOpen
  \bibfield  {author} {\bibinfo {author} {\bibfnamefont {G.~M.}\ \bibnamefont {Andolina}}, \bibinfo {author} {\bibfnamefont {M.}~\bibnamefont {Keck}}, \bibinfo {author} {\bibfnamefont {A.}~\bibnamefont {Mari}}, \bibinfo {author} {\bibfnamefont {M.}~\bibnamefont {Campisi}}, \bibinfo {author} {\bibfnamefont {V.}~\bibnamefont {Giovannetti}},\ and\ \bibinfo {author} {\bibfnamefont {M.}~\bibnamefont {Polini}},\ }\bibfield  {title} {\bibinfo {title} {Extractable work, the role of correlations, and asymptotic freedom in quantum batteries},\ }\href {https://doi.org/10.1103/PhysRevLett.122.047702} {\bibfield  {journal} {\bibinfo  {journal} {Phys. Rev. Lett.}\ }\textbf {\bibinfo {volume} {122}},\ \bibinfo {pages} {047702} (\bibinfo {year} {2019})}\BibitemShut {NoStop}%
\bibitem [{\citenamefont {Stahl}\ \emph {et~al.}(2024)\citenamefont {Stahl}, \citenamefont {Kewming}, \citenamefont {Goold}, \citenamefont {Hilder}, \citenamefont {Poschinger},\ and\ \citenamefont {Schmidt-Kaler}}]{Stahl2024}%
  \BibitemOpen
  \bibfield  {author} {\bibinfo {author} {\bibfnamefont {A.}~\bibnamefont {Stahl}}, \bibinfo {author} {\bibfnamefont {M.}~\bibnamefont {Kewming}}, \bibinfo {author} {\bibfnamefont {J.}~\bibnamefont {Goold}}, \bibinfo {author} {\bibfnamefont {J.}~\bibnamefont {Hilder}}, \bibinfo {author} {\bibfnamefont {U.~G.}\ \bibnamefont {Poschinger}},\ and\ \bibinfo {author} {\bibfnamefont {F.}~\bibnamefont {Schmidt-Kaler}},\ }\bibfield  {title} {\bibinfo {title} {Demonstration of energy extraction gain from non-classical correlations},\ }\href@noop {} {\bibfield  {journal} {\bibinfo  {journal} {arXiv:2404.14838}\ } (\bibinfo {year} {2024})}\BibitemShut {NoStop}%
\bibitem [{\citenamefont {Martinez}\ \emph {et~al.}(2016)\citenamefont {Martinez}, \citenamefont {Muschik}, \citenamefont {Schindler}, \citenamefont {Nigg}, \citenamefont {Erhard}, \citenamefont {Heyl}, \citenamefont {Hauke}, \citenamefont {Dalmonte}, \citenamefont {Monz}, \citenamefont {Zoller},\ and\ \citenamefont {Blatt}}]{Martinez2016}%
  \BibitemOpen
  \bibfield  {author} {\bibinfo {author} {\bibfnamefont {E.~A.}\ \bibnamefont {Martinez}}, \bibinfo {author} {\bibfnamefont {C.~A.}\ \bibnamefont {Muschik}}, \bibinfo {author} {\bibfnamefont {P.}~\bibnamefont {Schindler}}, \bibinfo {author} {\bibfnamefont {D.}~\bibnamefont {Nigg}}, \bibinfo {author} {\bibfnamefont {A.}~\bibnamefont {Erhard}}, \bibinfo {author} {\bibfnamefont {M.}~\bibnamefont {Heyl}}, \bibinfo {author} {\bibfnamefont {P.}~\bibnamefont {Hauke}}, \bibinfo {author} {\bibfnamefont {M.}~\bibnamefont {Dalmonte}}, \bibinfo {author} {\bibfnamefont {T.}~\bibnamefont {Monz}}, \bibinfo {author} {\bibfnamefont {P.}~\bibnamefont {Zoller}},\ and\ \bibinfo {author} {\bibfnamefont {R.}~\bibnamefont {Blatt}},\ }\bibfield  {title} {\bibinfo {title} {Real-time dynamics of lattice gauge theories with a few-qubit quantum computer},\ }\href {https://www.nature.com/articles/nature18318} {\bibfield  {journal} {\bibinfo  {journal} {Nat.}\ }\textbf {\bibinfo {volume} {534}},\ \bibinfo {pages} {516} (\bibinfo {year}
  {2016})}\BibitemShut {NoStop}%
\bibitem [{\citenamefont {Francica}\ \emph {et~al.}(2020)\citenamefont {Francica}, \citenamefont {Binder}, \citenamefont {Guarnieri}, \citenamefont {Mitchison}, \citenamefont {Goold},\ and\ \citenamefont {Plastina}}]{PhysRevLett.125.180603}%
  \BibitemOpen
  \bibfield  {author} {\bibinfo {author} {\bibfnamefont {G.}~\bibnamefont {Francica}}, \bibinfo {author} {\bibfnamefont {F.~C.}\ \bibnamefont {Binder}}, \bibinfo {author} {\bibfnamefont {G.}~\bibnamefont {Guarnieri}}, \bibinfo {author} {\bibfnamefont {M.~T.}\ \bibnamefont {Mitchison}}, \bibinfo {author} {\bibfnamefont {J.}~\bibnamefont {Goold}},\ and\ \bibinfo {author} {\bibfnamefont {F.}~\bibnamefont {Plastina}},\ }\bibfield  {title} {\bibinfo {title} {Quantum coherence and ergotropy},\ }\href {https://doi.org/10.1103/PhysRevLett.125.180603} {\bibfield  {journal} {\bibinfo  {journal} {Phys. Rev. Lett.}\ }\textbf {\bibinfo {volume} {125}},\ \bibinfo {pages} {180603} (\bibinfo {year} {2020})}\BibitemShut {NoStop}%
\bibitem [{\citenamefont {Francica}(2022)}]{PhysRevE.105.014101}%
  \BibitemOpen
  \bibfield  {author} {\bibinfo {author} {\bibfnamefont {G.}~\bibnamefont {Francica}},\ }\bibfield  {title} {\bibinfo {title} {Class of quasiprobability distributions of work with initial quantum coherence},\ }\href {https://doi.org/10.1103/PhysRevE.105.014101} {\bibfield  {journal} {\bibinfo  {journal} {Phys. Rev. E}\ }\textbf {\bibinfo {volume} {105}},\ \bibinfo {pages} {014101} (\bibinfo {year} {2022})}\BibitemShut {NoStop}%
\bibitem [{\citenamefont {Tacchino}\ \emph {et~al.}(2020)\citenamefont {Tacchino}, \citenamefont {Santos}, \citenamefont {Gerace}, \citenamefont {Campisi},\ and\ \citenamefont {Santos}}]{Tacchino2020}%
  \BibitemOpen
  \bibfield  {author} {\bibinfo {author} {\bibfnamefont {F.}~\bibnamefont {Tacchino}}, \bibinfo {author} {\bibfnamefont {T.~F.}\ \bibnamefont {Santos}}, \bibinfo {author} {\bibfnamefont {D.}~\bibnamefont {Gerace}}, \bibinfo {author} {\bibfnamefont {M.}~\bibnamefont {Campisi}},\ and\ \bibinfo {author} {\bibfnamefont {M.~F.}\ \bibnamefont {Santos}},\ }\bibfield  {title} {\bibinfo {title} {Charging a quantum battery via nonequilibrium heat current},\ }\href {https://doi.org/10.1103/PhysRevE.102.062133} {\bibfield  {journal} {\bibinfo  {journal} {Physical Review E}\ }\textbf {\bibinfo {volume} {102}},\ \bibinfo {pages} {1} (\bibinfo {year} {2020})}\BibitemShut {NoStop}%
\bibitem [{\citenamefont {Julia-Farre}\ \emph {et~al.}(2018)\citenamefont {Julia-Farre}, \citenamefont {Salamon}, \citenamefont {Riera}, \citenamefont {Bera},\ and\ \citenamefont {Lewenstein}}]{Julia}%
  \BibitemOpen
  \bibfield  {author} {\bibinfo {author} {\bibfnamefont {S.}~\bibnamefont {Julia-Farre}}, \bibinfo {author} {\bibfnamefont {T.}~\bibnamefont {Salamon}}, \bibinfo {author} {\bibfnamefont {A.}~\bibnamefont {Riera}}, \bibinfo {author} {\bibfnamefont {M.~N.}\ \bibnamefont {Bera}},\ and\ \bibinfo {author} {\bibfnamefont {M.}~\bibnamefont {Lewenstein}},\ }\bibfield  {title} {\bibinfo {title} {Bounds on the capacity and power of quantum batteries},\ }\bibfield  {journal} {\bibinfo  {journal} {Phys. Rev. Res.}\ }\href {https://doi.org/10.1103/PhysRevResearch.2.023113} {10.1103/PhysRevResearch.2.023113} (\bibinfo {year} {2018})\BibitemShut {NoStop}%
\bibitem [{\citenamefont {Son}\ \emph {et~al.}(2022)\citenamefont {Son}, \citenamefont {Talkner},\ and\ \citenamefont {Thingna}}]{Son2022}%
  \BibitemOpen
  \bibfield  {author} {\bibinfo {author} {\bibfnamefont {J.}~\bibnamefont {Son}}, \bibinfo {author} {\bibfnamefont {P.}~\bibnamefont {Talkner}},\ and\ \bibinfo {author} {\bibfnamefont {J.}~\bibnamefont {Thingna}},\ }\bibfield  {title} {\bibinfo {title} {Charging quantum batteries via otto machines: Influence of monitoring},\ }\href {https://doi.org/10.1103/PhysRevA.106.052202} {\bibfield  {journal} {\bibinfo  {journal} {Phys. Rev. A}\ }\textbf {\bibinfo {volume} {106}},\ \bibinfo {pages} {052202} (\bibinfo {year} {2022})}\BibitemShut {NoStop}%
\bibitem [{\citenamefont {Ahmadi}\ \emph {et~al.}(2024)\citenamefont {Ahmadi}, \citenamefont {Mazurek}, \citenamefont {Horodecki},\ and\ \citenamefont {Barzanjeh}}]{PhysRevLett.132.210402}%
  \BibitemOpen
  \bibfield  {author} {\bibinfo {author} {\bibfnamefont {B.}~\bibnamefont {Ahmadi}}, \bibinfo {author} {\bibfnamefont {P.}~\bibnamefont {Mazurek}}, \bibinfo {author} {\bibfnamefont {P.}~\bibnamefont {Horodecki}},\ and\ \bibinfo {author} {\bibfnamefont {S.}~\bibnamefont {Barzanjeh}},\ }\bibfield  {title} {\bibinfo {title} {Nonreciprocal quantum batteries},\ }\href {https://doi.org/10.1103/PhysRevLett.132.210402} {\bibfield  {journal} {\bibinfo  {journal} {Phys. Rev. Lett.}\ }\textbf {\bibinfo {volume} {132}},\ \bibinfo {pages} {210402} (\bibinfo {year} {2024})}\BibitemShut {NoStop}%
\bibitem [{\citenamefont {Gyhm}\ \emph {et~al.}(2022)\citenamefont {Gyhm}, \citenamefont {\ifmmode~\check{S}\else \v{S}\fi{}afr\'anek},\ and\ \citenamefont {Rosa}}]{PhysRevLett.128.140501}%
  \BibitemOpen
  \bibfield  {author} {\bibinfo {author} {\bibfnamefont {J.-Y.}\ \bibnamefont {Gyhm}}, \bibinfo {author} {\bibfnamefont {D.}~\bibnamefont {\ifmmode~\check{S}\else \v{S}\fi{}afr\'anek}},\ and\ \bibinfo {author} {\bibfnamefont {D.}~\bibnamefont {Rosa}},\ }\bibfield  {title} {\bibinfo {title} {Quantum charging advantage cannot be extensive without global operations},\ }\href {https://doi.org/10.1103/PhysRevLett.128.140501} {\bibfield  {journal} {\bibinfo  {journal} {Phys. Rev. Lett.}\ }\textbf {\bibinfo {volume} {128}},\ \bibinfo {pages} {140501} (\bibinfo {year} {2022})}\BibitemShut {NoStop}%
\end{thebibliography}%

\end{document}